\documentclass[useAMS,usenatbib]{mn2e}
\usepackage{epsfig,rotate,graphicx}
\usepackage[fleqn]{amsmath}
\usepackage{amssymb}
\usepackage{array}
\usepackage{subfigure}
\usepackage{lscape}
\usepackage{bm}

\newcommand{\nat}{{ Nature, }}

\newcommand{\apjs}{{ ApJS, }}

\newcommand{\mnras}{{ MNRAS, }}

\newcommand{\aaps}{{ A\&AS, }}

\newcommand{\apj}{ApJ}
\newcommand{\aap}{A\&A}
\newcommand{\apss}{APSS}
\newcommand{\apjl}{ApJL}
\newcommand{\araa}{ARAA}

\newcommand{\ciso}{c_\mathrm{iso}}
\newcommand{\avg}[1]{\langle #1 \rangle}
\newcommand{\totg}{\Gamma_\mathrm{tot}}
\newcommand{\fargo}{\texttt{FARGO }}
\newcommand{\zeus}{\texttt{ZEUS }}

\defcitealias{lin12}{LP12}

\title[Orbital migration induced by unstable gaps]{Orbital migration
  of giant planets induced by gravitationally unstable gaps: the effect of
  planet mass}   

\author[Cloutier and Lin]{ Ryan Cloutier 
  \thanks{E-mail: cloutier@cita.utoronto.ca} and Min-Kai Lin 
  \thanks{E-mail: mklin924@cita.utoronto.ca} \\ 
  Canadian Institute for Theoretical Astrophysics, 
  60 St. George Street, Toronto, ON, M5S 3H8, Canada \\
}

\begin{document}

\maketitle

\begin{abstract}
It has been established that self-gravitating disc-satellite 
interaction can lead to the formation of a gravitationally unstable
gap. Such an instability may significantly affect the orbital migration
of gap-opening perturbers in self-gravitating discs. In this paper, we
extend the two-dimensional hydrodynamic simulations of
\citeauthor{lin12} to investigate the role of the perturber or planet
mass on the gravitational stability of gaps and its impact on orbital
migration. We consider giant planets with planet-to-star mass ratio 
$q\equiv M_{p}/M_{*}\in[0.3,\,3.0]\times10^{-3}$ (so that $q=10^{-3}$
corresponds to a Jupiter mass planet if $M_*=M_{\sun}$),  
in a self-gravitating disc with disc-to-star mass ratio
$M_d/M_*=0.08$, aspect ratio $h=0.05$, and Keplerian Toomre parameter
$Q_{k0}=1.5$ at 2.5 times the planet's initial orbital radius. These
planet masses correspond to $\tilde{q}\in[0.9,\,1.7]$, where
$\tilde{q}$ is the ratio of the planet Hill radius to 
the local disc scale-height. 
Fixed-orbit simulations show that all planet masses we
consider open gravitationally unstable gaps, but the instability is
stronger and develops sooner with increasing planet
mass. The disc-on-planet torques typically become more positive with
increasing planet mass. In freely-migrating simulations, we observe
faster outward migration with increasing planet mass, but only for
planet masses capable of opening unstable gaps early on. For
$q=0.0003$ ($\tilde{q}=0.9$), the planet undergoes rapid inward type III migration before
it can open a gap. For $q=0.0013$ ($\tilde{q}=1.5$) we find it is possible to balance
the tendency for inward migration by the positive torques due to an 
unstable gap, but only for a few 10's of orbital periods. We find 
the unstable outer gap edge can trigger outward type III migration,
sending the planet to twice it's initial orbital radius on dynamical
timescales. We briefly discuss the importance of our results in the context of  
giant planet formation on wide orbits through disc fragmentation. 

\end{abstract}


\section{Introduction}
From the discovery of `hot Jupiters' \citep[e.g. 51 Peg b,][]{mayor95} to
long-period giant planets \citep[e.g. HR 8799b,][]{marios08}, the wide
range of observed exoplanet orbital radii suggest that orbital migration due to
gaseous disc-satellite interaction may play an important role in
planet formation theory. Since its initial development
\citep{goldreich79,goldreich80}, disc-satellite
interaction has been studied with the inclusion of increasingly
complex physics. For a review of the theory and recent advancements,
see \cite{kley12} and \cite{baruteau13}.    

A less well-explored area is the interaction between a planet and 
large-scale instabilities in the disc. For example, recent works have
shown that disc gaps induced by a planet can be dynamically
unstable under appropriate conditions
\citep{koller03,li05,lin10,lin11a}. Gap-opening requires
a sufficiently massive planet \citep{lin86} and/or low
viscosity disc \citep{rafikov02, dong11, duffell13}.


In a self-gravitating disc, planet gaps  may become 
gravitationally unstable \citep{meschiari08,lin12b} even if the
initial disc is Toomre stable \citep{toomre64}. The result of this
instability is the development of large-scale spiral arms associated
with the gap edge, which exert significant torques on the planet
\citep{lin11b}. This gravitational edge instability, and its impact on
orbital migration, has been less appreciated. However, it may be
relevant to planet formation theories requiring a self-gravitating disc
\citep[e.g.,][]{boss97,naya10,naya13}. We remark that in the `tidal
downsizing' theory, \citeauthor{naya10} discusses gap-opening in
massive discs, a situation that we consider in this work. 
 
In a previous study, \citet[hereafter \citetalias{lin12}]{lin12}
simulated the orbital migration of a gap-opening giant planet in
self-gravitating discs which became gravitationally unstable only in the
presence of a planet gap. They found a gravitationally unstable outer
gap edge induced \emph{outward} orbital migration. 


\citetalias{lin12} fixed the planet mass in their
simulations. However, since the instability is associated with the 
gap, and the gap structure depends on the planet mass, we expect the
gravitational stability of planet gaps to also depend on the planet 
mass. The present study is a natural follow-up to \citetalias{lin12} in
which we investigate the role of planet mass.  


This paper is organized as follows. In the next subsection, we 
review the basic properties of the gravitational instability
associated with planet gaps, and further explain the motivation
for our study. We describe our disc-planet models and numerical
methods in \S\ref{model}. We present results for fixed-orbit
simulations in \S\ref{sectfix} and freely-migrating simulations in
\S\ref{sectmig}. We show it is possible to for edge modes to
counter-act inward migration, but not indefinitely. We find the 
unstable gap edge easily triggers rapid outward migration. An 
example of this phenomenon is discussed in \S\ref{fiducial}. We summarize in \S\ref{summary}
with a discussion of possible applications of our results to giant
planets on wide orbits, and important caveats of our models. 

\subsection{Gravitational instability of planet gaps}\label{overview}

Gaps induced by giant planets are associated with extrema in the disc
potential vorticity (PV) profile  
$\eta=\kappa^2/2\Omega\Sigma$, where $\kappa$ is the epicycle frequency, $\Omega$ is the rotation rate and $\Sigma$ is the surface density. 
It is well-known that the presence of PV extrema permits dynamical instability \citep[e.g.][]{papaloizou85, li00, lin10}.
In the case of a planet gap, local $\mathrm{max}(\eta)$ and local $\mathrm{min}(\eta)$  results from 
PV generation and destruction across spiral shocks induced by the planet \citep{koller03,li05,lin10}. 

The PV profile of a planet gap resembles its Toomre parameter profile  
$Q=c_s\kappa/\pi G \Sigma$ \citep{lin11b}, where $c_s$ is the sound-speed 
and $G$ is the gravitational constant.  
Thus, we may also associate planetary gap instabilities  
with $\mathrm{max}(Q)$ or $\mathrm{min}(Q)$. 

\begin{figure}
  \centering
  \includegraphics[width=\linewidth]{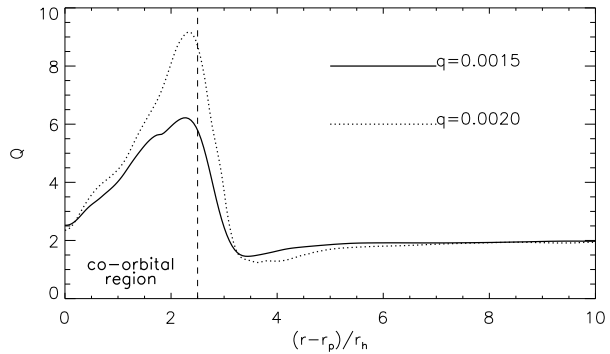}
  \caption{Azimuthally averaged Toomre parameter for gaps
opened by planets with different masses ($q$ is the planet-to-star mass ratio).  
The gap edge spiral instability is 
    associated with the local $\mathrm{max}(Q)$, situated at
    approximately 2 Hill radii from the planet's orbital radius. This
    is to be compared with the co-orbital region of a giant planet,
    which is approximately $|r-r_p|\lesssim 2.5r_h$ (vertical line). So the instability
    is associated with a feature \emph{just inside} the gap. \label{profiles}}
\end{figure}

Fig. \ref{profiles} shows typical $Q$-profiles for gaps induced by
giant planets. Only the outer disc is shown since
this is where instability is most prominent in our models. The horizontal axis is plotted in units of 
Hill radii $r_h$ away from the planet's orbital radius (defined later). The co-orbital region for massive planets, 
within which fluid particles execute horseshoe turns upon encountering the planet, is 
approximately within $2.5r_h$ of the planet's orbital radius \citep{artymowicz04b,paardekooper09,lin10}. 
So the local $\mathrm{max}(Q)$ is located just inside the co-orbital region, while the
local $\mathrm{min}(Q)$ is located just outside.  

In a self-gravitating disc, \cite{lin11b} showed that 
there is a gravitational instability associated with the local $Q$-\emph{maximum} at
a planetary gap edge\footnote{See also \cite{meschiari08}, who
originally suggested gravitational instability of planetary gaps based on
prescribed disc profiles.}. By association we mean that the co-rotation radius $r_c$ of the unstable mode coincides with or is close to that of the PV or $Q$ maximum. 
The local $\mathrm{max}(Q)$ is situated just inside the gap edge
(Fig. \ref{profiles}). Thus, the instability 
presents non-axisymmetric disturbances in the planet's co-orbital region.
In their disc models where the instability is associated with the
outer gap edge, \citetalias{lin12} found the passage of an associated spiral arm leads to a 
net positive torque applied to the planet, because the spiral arm  
supplies material to execute inward horseshoe turns upstream of the planet. 
This interaction is explicitly illustrated in Fig. 8 of \citetalias{lin12}.

Fig. \ref{profiles} shows that the gap profile is a
function of planet mass. It is not obvious how this 
affects the positive torques due to the gap gravitational instability
described above. Increasing the planet mass makes the gap edges sharper,
which should favour instability and increase spiral mode amplitudes, and lead to stronger torques.   
On the other hand, a larger planet mass opens a 
deeper gap with lower surface density (reflected by the larger
$Q$-value in Fig. \ref{profiles}). The latter effect should decrease the magnitude of
disc-on-planet torques originating from the co-orbital region. 

The purpose of this paper is to clarify, through
hydrodynamical numerical experiments, the role of 
planet mass on the gravitational stability of gaps and the subsequent 
orbital migration due to the interaction with the instability. 
\cite{lin11b} found that Saturnian mass planets ultimately migrated
inward, despite brief phases of outward migration induced by the gap 
edge instability. However, \citetalias{lin12} simulated 2-Jupiter mass
planets and found sustained outward migration. It is of
interest to examine the possibility suggested in \citetalias{lin12}: zero net migration due to a 
balance between the outward migration induced by the instability and the 
tendency for inward type II migration \citep{lin86}. 

\section{Disc-planet model}\label{model}
We consider a two-dimensional (2D) gaseous self-gravitating protoplanetary disc of
mass $M_{d}$ orbiting a central star of mass $M_{*}$. 
Embedded within
the disc is a planet of mass $M_{p}$. 
We use $(r,\phi)$  
plane polar co-ordinates centered on the star. The disc-planet model is
the same as that in \citetalias{lin12}, but we describe it here for
ease of reference. We adopt units such that $G=M_*=1$. 

The disc has radial extent $r\in [r_{i},r_{o}]=[1,25]$. The
strength of disc self-gravity is characterized by specifying 
$Q_{k0}\equiv Q_k(r_o)$, 
where  
\begin{align}\label{Qk}
  Q_k(r)\equiv\frac{\ciso\Omega_k}{\pi G\Sigma}
\end{align}
is the Keplerian Toomre parameter for thin discs. In Eq. \ref{Qk},  
$\ciso$ is the isothermal sound-speed defined
below, $\Omega_{k}(r)=\sqrt{GM_{*}/r^{3}}$ is the Keplerian orbital
frequency. 
For all our simulations, the disc surface density is initialized with
$\Sigma(r)\propto r^{-3/2}$ for $r\gg r_i$, such that
$Q_{k0}=1.5$.  Our disc model is gravitationally stable to 
  axisymmetric  perturbations according to the Toomre criterion
  \citep{toomre64}. 

It should be emphasised that the outcome of the simulations 
  depends on the initial $Q_{k0}$. Our choice of $Q_{k0}=1.5$
  favours the instability-induced outward migration as seen in
  \citetalias{lin12} (where outward migration slows down for higher $Q_{k0}$, and is not observed
  within their simulation timescale of $100P_0$ for $Q_{k0}=2$).  
  The surface density normalization $Q_{k0}=1.5$ gives a disc mass of $M_d=0.08M_*$. 
  This is $\sim8$ times larger than the traditional minimum-mass Solar
  nebula \citep[MMSN, ][]{weiden77}, but comparable to the more massive MMSN
  constructed by \cite{desch07}.


We adopt a locally isothermal equation of state so that the 
vertically integrated pressure $p=c_s^2\Sigma$. Without the planet the 
sound-speed $c_s=\ciso\equiv h r \Omega_k$, where $h$ is 
the disc aspect-ratio. We fix $h=0.05$. The sound-speed is modified
close to the planet when it is introduced (see \S\ref{planet_config}).  
We also impose a constant kinematic viscosity $\nu=10^{-5}r_i^2\Omega_k(r_i)$.
This corresponds to an alpha-viscosity of order $10^{-3}$, which is
typical for disc-planet simulations.  

\subsection{Planet configuration}
The main parameter that we vary is the planet mass $M_p\equiv
qM_*$, where $q$ is the planet-to-star mass ratio. We consider
$q\in[0.3,3]\times10^{-3}$, but will be primarily interested in cases
with $q\simeq 10^{-3}$. If $M_*=M_{\sun}$ then $q=0.001$ corresponds 
to a Jupiter-mass planet and $q=0.0003$ corresponds to a Saturn-mass
planet. The position of the planet is denoted
$\bm{r}_p=(r_p,\phi_p)$. The planet is introduced on a
  circular orbit of radius $r_{p0}= r_{p}(t=20P_0)=10$ where
  $P_0=2\pi/\Omega_k(r_{p0})$. This corresponds to $Q_k(r_{p0})=2.77$,
  and the mass within $|r-r_{p0}|\lesssim 2.5r_h$ is initially $\simeq
  10M_p$ for $q=10^{-3}$.

The planet mass is 
  ramped up from zero to its full value over $10P_0$. Thus the planet is 
  fully introduced into the disc by $t=30P_0$. Orbital migration is
  allowed for $t>30P_0$, if at all. The planet's gravitational  
potential is softened with
a softening length $\epsilon_p=0.6H$. Accretion onto the planet is 
neglected, but since we compute the full self-gravity of
the gas, material gravitationally bound to the planet effectively
increases its mass. 

\subsubsection{Gap opening criteria}
\cite{crida06} 
showed that gap opening by a planet depends specifically on its mass, the disc scale-height $H=hr$, and the viscosity of the disc. Their criterion for gap opening as a function of $q$, $H$, and $\nu$ is

\begin{equation}
\frac{3}{4} \frac{H}{r_{h}} + \frac{50}{q} \left(\frac{\nu}{r_{p}^{2} \Omega_{p}}\right) \lesssim 1, \label{criteria}
\end{equation}
where 
$r_{h}=(q/3)^{1/3} r_p$ is the Hill radius of the planet and $\Omega_p\equiv\Omega(r_p)$. 

This criteria is useful in determining which of our parameter survey values $q$, are able to induce a gap in the disc. 
Specifically, we can solve for the critical gap-opening mass, $q_{c}$ for which the left-hand side of Eq. \ref{criteria} equals unity. 
Doing so, we find that $q_{c}=5 \times 10^{-4}$. We therefore expect all planet masses with $q>q_{c}$ to open gaps. The case with $q=3\times10^{-4}$ is somewhat smaller than $q_c$, but we will find even a partial gap 
can be gravitationally unstable. 
  
\subsubsection{Equation of state (EOS)}\label{planet_config}
In the presence of the planet, we adopt the following prescription for
the sound-speed:  
\begin{equation}\label{peplinski_eos}
  c_{s}=\frac{\ciso
    h_{p}d_{p}}{[(hr)^{7/2}+(h_{p}d_{p})^{7/2}]^{2/7}}\left(1+\frac{\Omega_{kp}^{2}}{\Omega_k^2}\right)^{1/2},     
\end{equation} 
where $\Omega_{kp}^2= GM_{p}/d_{p}^{3}$ with
$d_{p}^{2}=|\mathbf{r}-\mathbf{r}_{p}|^{2}+\epsilon_p^{2}$ and 
$h_{p}$ is a dimensionless parameter. Note that $c_s\to\ciso$ as
$d_p\to\infty$. 

Eq. \ref{peplinski_eos} is taken from \cite{peplinski08a}, and is used
here to increase the disc temperature near the planet relative to
$\ciso$. The magnitude of this increase is controlled by $h_p$. This 
temperature increase mitigates accumulation of gas near 
$\bm{r} = \bm{r}_p$. This would occur if we set $c_s=\ciso$ (implying
the disc temperature is unaffected by the planet), which may lead to
spurious torques arising from gas near the planet due to the 
diverging potential and limited resolution. 

Physically, we expect gas near the planet to heat up as it falls
into the planet potential. 
The appropriate value for $h_p$ depends on detailed thermodynamics
occurring near the planet, which would depend on planet mass. 
However, since use of this EOS in the present study is motivated by numerical
considerations, we simply choose $h_p$ to ensure $c_s/\ciso > 1$ 
everywhere. In practice, we choose $h_p=0.5$ for all planet masses except
for $q=3\times10^{-4}$, for which $h_p=0.65$ was needed. 

\subsection{Numerical simulations}\label{numerics}
We evolve the disc-planet system using the \fargo code \citep{masset00a,baruteau08}. 
\fargo solves the 2D hydrodynamic equations using a finite-difference scheme
similar to the \zeus code \citep{stone92}, except with a modified
azimuthal transport algorithm which circumvents the time-step limitation set by the 
inner disc boundary. The self-gravity solver is described in \cite{baruteau08}. When allowed to
respond to disc forces, the planet's motion is integrated with a
fifth-order Runge-Kutta scheme. Indirect potentials are included
  to account for the non-inertial reference frame. The disc indirect
  potential is not expected to play a significant role because our
  discs are not very massive \citep[cf.][]{adams89,shu90,kratter10b}.  

The disc is divided in to $N_r\times N_\phi$ cells in radius and azimuth, respectively. 
The radial grid is logarithmically spaced while the azimuthal grid is uniformly spaced. 
We use $(N_r,N_\phi)=(512,1024)$ for fixed-orbit simulations (\S \ref{sectfix}) and $(N_r,N_\phi)=(1024,2048)$
for simulations where the planet is allowed to migrate (\S \ref{sectmig}). 
In the latter case, the resolution is increased in order to resolve regions close to the planet,
where co-orbital torques arise and were found to be responsible for the outward migration seen in \citetalias{lin12}.

We then apply open boundaries in the radial direction and periodic boundary conditions
in azimuth. 

We initialize the disc azimuthal velocity $v_\phi$  from centrifugal
balance with stellar gravity, self-gravity and pressure forces.  
The initial radial velocity is $v_r = 3\nu/r$.


\section{Fixed Orbit Simulations} \label{sectfix}

We first examine how the gap edge spiral instability (`edge modes' hereafter) and 
the associated disc-planet torques depend on planet mass. To focus
on these issues we neglect orbital migration and 
hold the planet on a fixed circular orbit throughout the simulations
presented in this section. 

It is important to keep in mind that fixed-orbit simulations 
suppress disc-planet torques \emph{due} to orbital migration. Such
torques can be expected for giant planets in massive discs 
undergoing type III migration
\citep{masset03,peplinski08a}. Nevertheless, we find these numerical
experiments  useful to aid the interpretation of
freely-migrating cases considered later.  

\subsection{Gap evolution} \label{gapsect}  
Edge modes are associated with local PV maxima 
located just inside the gap (see
Fig. \ref{profiles}). 
The presence of such instabilities will
therefore be reflected in the gap properties such as gap depth. This
is a useful quantity to examine because edge modes have a
`gap-filling' effect. This signifies material being brought into the 
co-orbital region of the planet, which can subsequently provide a 
torque \citepalias{lin12}.  

In our disc models the Toomre stability parameter decreases
with radius, which favours instability of the outer gap edge rather 
than the inner gap edge. We therefore focus on the gap structure in
$r>r_p$.     

Following \citetalias{lin12}, we define the radius of the outer gap edge
$r_\mathrm{rout}>r_p$ such that $\delta \Sigma(r_\mathrm{out})=0$,
where 
\begin{equation}
  \delta \Sigma (r) = \bigg \langle
  \frac{\Sigma-\Sigma(t=0)}{\Sigma(t=0)} \bigg
  \rangle_{\phi} \label{pert} 
\end{equation}      
is the one-dimensional relative surface density perturbation and
$\avg{\cdot}_\phi$ denotes azimuthal average.  The outer gap depth is   
defined to be the average value of $\delta\Sigma$ in
$r\in[r_p,r_\mathrm{out}]$ and is denoted $\langle \delta \Sigma
\rangle |^{r_\mathrm{out}}_{r_{p}}$. This is a negative quantity
because the gap is a surface density deficit, but for convenience 
we will refer to the magnitude of the outer gap depth simply as the
gap depth.  

\begin{figure}
  \centering
  \includegraphics[scale=0.6]{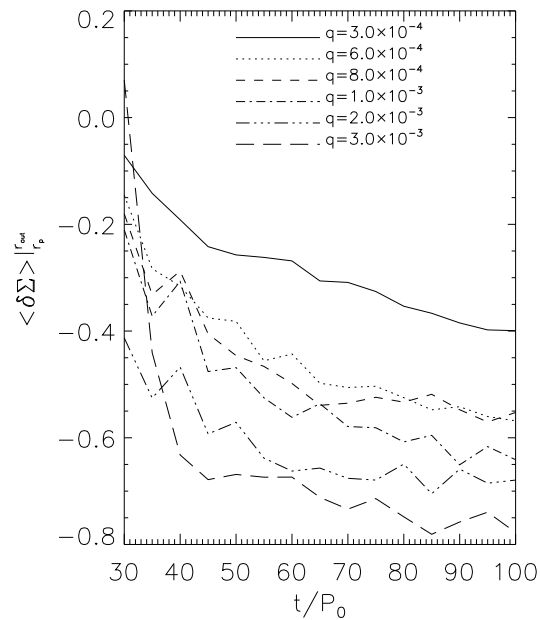}
  \caption{Time evolution of the outer gap depth as a
    function of planet mass. The gap becomes deeper with time, but the evolution is
    non-smooth due to the development of edge modes associated with
    the gap edge. The case with $q=0.003$ underwent fragmentation, 
  which resulted in a more shallow gap at $t=30P_0$. \label{gap}}  
\end{figure}

Fig. \ref{gap} shows the evolution of the gap depth for a range of 
planet masses $q\in[0.3,3.0]\times10^{-3}$. Increasing
$q$ results in deeper gaps. Local self-gravity is
expected to be less important for deeper gaps 
because the surface density is relatively low compared to $\Sigma(t=0)$. 
This should 
discourage gravitational instabilities. 
However, increasing
$q$ produces steeper gap edges which are expected to favour edge
modes. In fact, Fig. \ref{gap} indicates the latter effect is more
important, as discussed below. 

\begin{figure}
\centering
\includegraphics[scale=0.39]{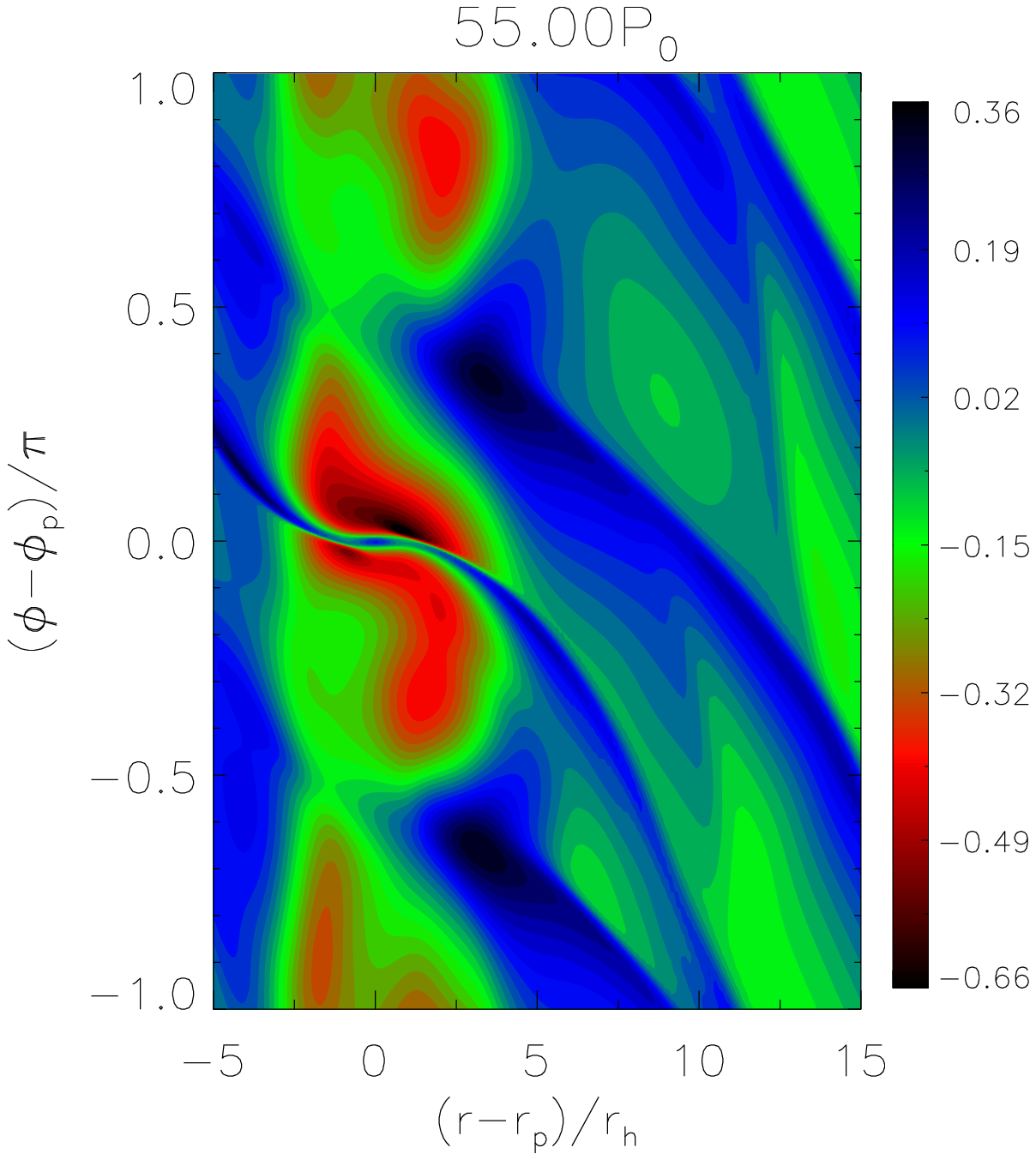}\includegraphics[scale=0.39,clip=true,trim=2.21cm
  0cm 0cm 0cm]{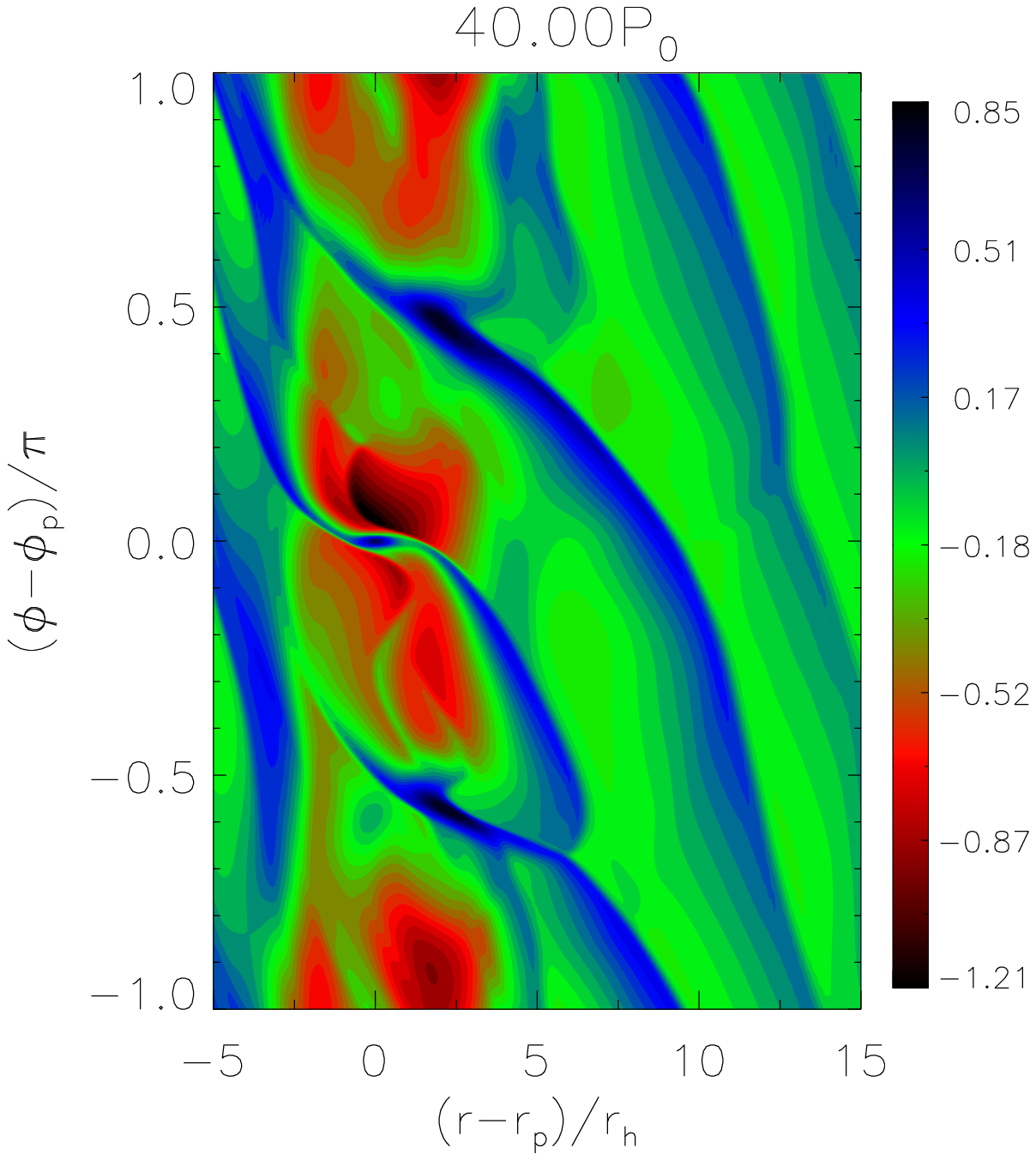}
\caption{Gravitational instability associated with the gap opened
  by a Saturnian mass planet ($q=3\times10^{-4}$, left) and a Jovian
  mass planet ($q=10^{-3}$, right). The logarithmic surface density
  perturbation $\log{[\Sigma/\Sigma(t=0)]}$ is shown. Note that edge
  modes develop earlier with increasing $q$.  
  \label{edge}} 
\end{figure}

An important feature in Fig. \ref{gap} is that the gap depth does not
decrease monotonically. This signifies instability. 
In some cases a `spike' appears in the evolution where the gap fills
up temporarily, which is attributed to the protrusion of edge mode
spiral arms into the gap \citepalias[][see also Fig. \ref{edge}]{lin12}. Thus the rate of
gap-opening can be slowed down by edge modes. This is explicitly shown
in Fig. \ref{gap}. As a result, the gap is not completely cleared
even for $q=0.002$  ($\delta\Sigma\sim -0.7$) which would be expected
if no instabilities develop.   

We find edge modes set in earlier with increasing planet mass.    
The spiral arms develop at $t\sim 50P_0$ for $q=3\times10^{-4}$, which
is consistent with previous work \citep{lin11b}. For
$q\in[0.6,2.0]\times10^{-3}$ the spiral arms develop at $t\sim40P_0$.
An example of the edge modes is shown in Fig. \ref{edge}. 

The instability is also stronger with increasing $q$. This is
reflected in Fig. \ref{gap} by the steep increase in $\langle \delta
\Sigma \rangle |^{r_\mathrm{out}}_{r_{p}}$ at $t\simeq40P_0$ for 
$10^3q=0.8,\,1.0,\,$and 2.0 (the `spikes'). By contrast, such gap-filling effect is
only just noticeable for $q=3\times10^{-4}$ with a slight
decrease in the rate at which the gap deepens around $t\sim
50P_0$ (the small bump), i.e. when instability sets in.

We observed fragmentation in the run with $q=0.003$. In this case 
the spiral wake induced by the planet undergoes
fragmentation. There is no well-defined gap and 
and large-scale coherent edge mode spirals were 
not identified (cf. Fig. \ref{edge}). 
Planet-induced fragmentation of
self-gravitating discs have been previously examined by
\cite{armitage99} and \cite{lufkin04}. We do not consider this fragmenting
case further for fixed-orbit or migration simulations. 


\subsection{Torque measurements} \label{tor}
In this section we compare the disc-on-planet torques  
$\totg$ measured from the above simulations. 
Here, we also apply an exponential envelope to taper 
off torque contributions from within the planet's Hill sphere. This
avoids potential numerical artifacts from this region because of the
low-resolution adopted, and 
allows us to focus on the effect of edge mode spiral arms, which 
typically reside in $r\gtrsim r_p+2r_h$ \citepalias{lin12}. 

\begin{figure}
\centering
\includegraphics[scale=0.4,clip=true,trim=1cm 2cm 0.8cm 0cm]{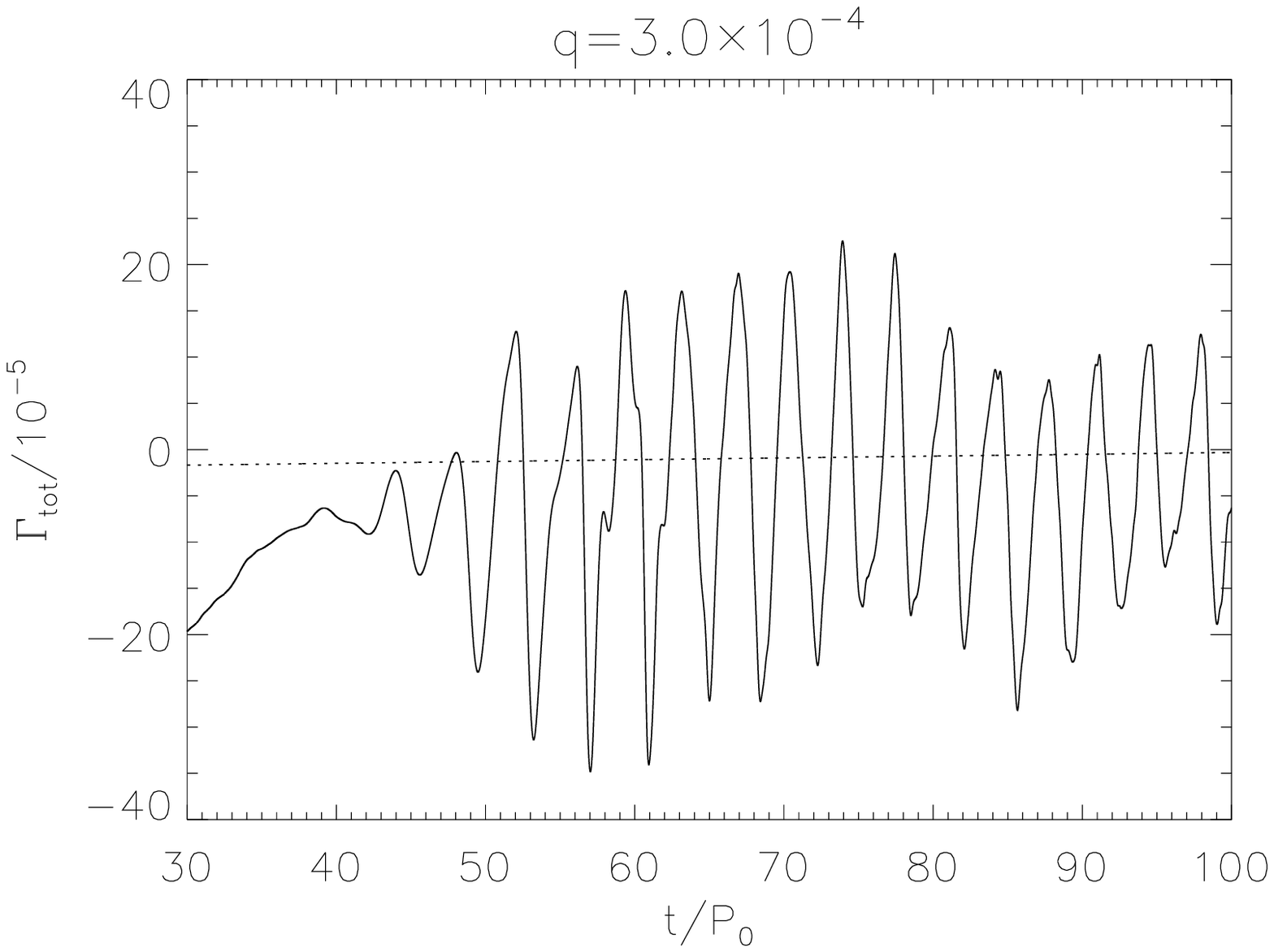} 
\includegraphics[scale=0.4,clip=true,trim=0.6cm 0cm 0.5cm 0cm]{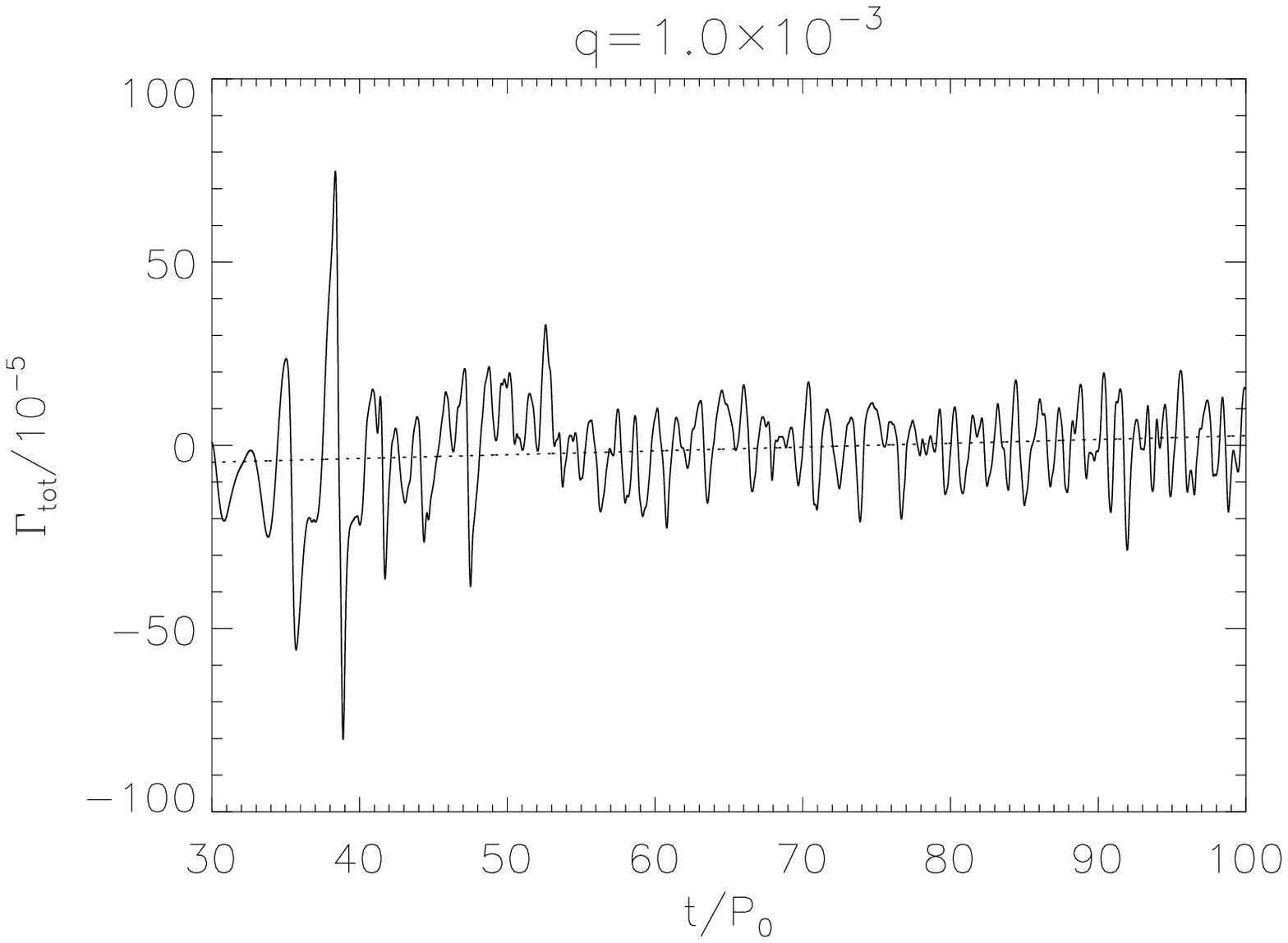}
\caption{Instantaneous specific disc-on-planet torques, in code units, 
  for $q=3.0\times10^{-4}$ (top) and $q=10^{-3}$ 
  (bottom). The planet is held on a fixed circular orbit. 
  The straight line in each plot represents a linear
  fit which indicates the time rate of change in the total torque.  
Large oscillations signify the development of edge modes,
  which sets in earlier with increasing $q$. 
  \label{torque}} 
\end{figure}

The previous section showed that edge modes develop earlier and are
stronger with Jovian planetary gaps than sub-Jovian gaps. 
Fig. \ref{torque} compares the evolution of the total torque
between these two cases.     
Oscillatory torques signify the presence of edge modes, 
which occurs almost immediately for 
$q=10^{-3}$, but takes longer for $q=3\times10^{-4}$. 
We also find the average amplitude of the torque oscillations 
to be larger with increasing planet mass as a consequence of
stronger instability. This is despite the fact that increasing $q$
lowers the average surface density in the gap region.

For $q=3\times10^{-4}$, the torque is at first negative 
due to the planetary wakes (or differential Lindblad torque), until
the point at which edge modes develop ($t \sim 50P_{0}$). The contrast between 
the torque evolution before and after the onset of instability  in 
 Fig. \ref{torque} shows that edge modes significantly modify 
disc-planet torques. That is, the development of edge modes may render the
Lindblad torques 
negligible compared to those provided by the edge
mode spirals. 

Fig. \ref{torque} also shows that the amplitude of the torque oscillations is 
largest when the instability first sets in (it saturates later).   
This suggests that in their initial stages of 
development, edge mode spirals can give the planet a `kick', which can 
influence the nature of the subsequent orbital migration. 
However, instantaneous torques in the presence of edge
modes may be positive or negative, so it is difficult to predict the 
direction of migration based solely on Fig. \ref{torque}. 

It takes $\sim 10P_0$ longer for the $q=3\times10^{-4}$ gap to become
unstable than the $q=10^{-3}$ gap. Although this is a small difference, for 
giant planets in massive discs considered in this
paper, type III migration may be applicable \citep{masset03}, 
but is suppressed in fixed-orbit simulations. 
The timescale for type III orbital migration 
is on the order a few 10's of $P_0$ \citep{peplinski08b}. 
This suggests that edge modes may be irrelevant if they do not 
develop early on, because type III migration could occur before edge
modes develop. We confirm this for sub-Jovian planets in \S \ref{sectmig}. 

\subsubsection{Time-averaged torques} 
\begin{figure}
\centering
\includegraphics[scale=0.5,clip=true,trim=1cm 0cm 0cm 1cm]{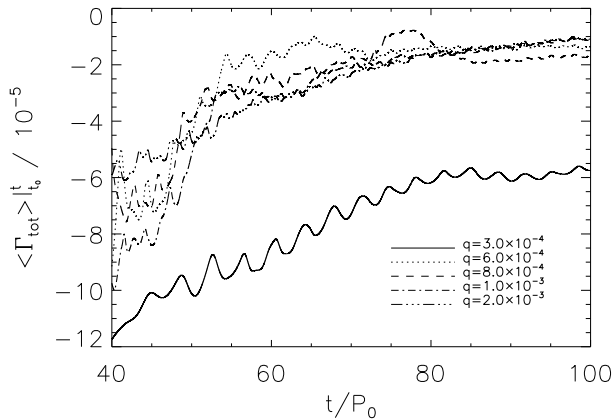}
\caption{Running time-average specific disc-on-planet torque for a range of
  planet masses when the gap is unstable to edge modes.  
  \label{timetor}} 
\end{figure}

Here, we compute time-averaged torques in order to
remove the oscillatory behaviour observed above and compare
disc-planet torques for a range of $q$ with 
unstable gaps. We define the running time-averaged torque as 
\begin{align}
  \langle \Gamma_\mathrm{tot} \rangle |^{t}_{t_{0}} \equiv 
  \frac{1}{t-t_{0}} \int^{t}_{t_{0}} \Gamma_\mathrm{tot}(t^\prime)
  dt^\prime, 
\end{align}
and we set $t_0=30P_0$. We plot the running time-averaged torque evolution in
Fig. \ref{timetor} , while Table \ref{avgs} compares the values 
of average torques at the end of the simulation. 

Fig. \ref{timetor} shows that edge modes cause the average torque to
become more positive with time. 
The case with $q=3\times10^{-4}$ has noticeably more negative torque values than 
cases with larger $q$.  
Increasing the planet mass, and hence the instability strength, typically 
results in more positive disc-on-planet torques (see Table
\ref{avgs}). However, the trend is not a clean function of $q$. This
may be due to the fact that increasing planet mass also increases the
magnitude of the differential Lindblad torque, which is negative. 
  
\begin{table}
  \centering
  \begin{tabular}[h]{ccc} 
    $10^3q$ & $10^5\langle \Gamma_\mathrm{tot} \rangle$ \\
    \hline \hline
    0.3 & -5.75 \\
    0.6 & -1.37 \\
    0.8 & -1.66 \\
    1.0 & -1.07 \\
    2.0 & -1.09 \\

    \hline 
  \end{tabular}
  \caption{The average specific torque acting on each planetary mass (in
    code units) for fixed-orbit simulations. 
    \label{avgs}} 
\end{table}

\subsection{Implications from fixed-orbit simulations} 
The above numerical experiments show that for the adopted disc
parameters, gaps opened by giant planets in self-gravitating discs are
unstable to edge modes even for a Saturnian-mass planet ($q=3\times10^{-4}$), which is
expected to open a partial gap at most (over sufficiently long timescales).   
However, edge modes develop later with decreasing $q$. 

The overall
effect of increasing $q$ is to increase the strength of the edge mode  
instability. This is despite the increased tidal torque 
lowering the gap surface density, an effect that should
discourage gravitational instabilities. This effect is, however,
outweighed by the increased sharpness of gap edges due to increased
$q$, which favour edge modes.  

Given that the edge mode-modified torques oscillate
rapidly between positive and negative values, the simulations here
cannot be used to predict the direction of orbital migration in the
presence of edge modes. Nevertheless, we expect inwards migration to
be become less favourable with increasing $q$ based on time-averaged
torques, which display a more positive torque with increasing
planet mass.

\section{Simulations with a freely migrating planet} \label{sectmig}  
In this second set of calculations, we allow the planet to respond to
the disc forces after its potential is fully 
introduced. These simulations permit type III migration, for which the 
torque results from fluid crossing the planet's orbital radius and the
torque magnitude is proportional to the migration rate
\citep{masset03}.   
Such torques originate close to the planet. Accordingly we double the
grid resolution in both directions from the previous experiments
(\S\ref{numerics}). 

We examine 
cases with $10^3q = 0.3,\,0.8,\,1.0,\,1.3,\,1.5$ and 2.0. Our 
specific aim is to see whether or not it is possible to balance the
tendency for inward migration with the positive torques induced by the
edge modes.   

We find the resulting migration is broadly consistent with torque
measurements made above. In particular, only planet masses that 
induce strong instability during gap formation experience the effect of
positive torques brought upon them by edge modes. Otherwise, the 
planet falls in on dynamical timescales. 

In fact, only for the cases $q\geq1.3\times10^{-3}$ did we find the
situation originally envisioned for our study: a planet residing in a
gap with large-scale edge mode spiral arms at the edge over long
($\sim 100 P_{0}$) timescales (such as
Fig. \ref{edge}). We will focus on such cases later, but first briefly
review the smaller planet mass runs.    
 
\subsection{Inward migration of sub-Jovian mass
  planets} \label{inward} 
\begin{figure}
\centering
\includegraphics[width=\linewidth]{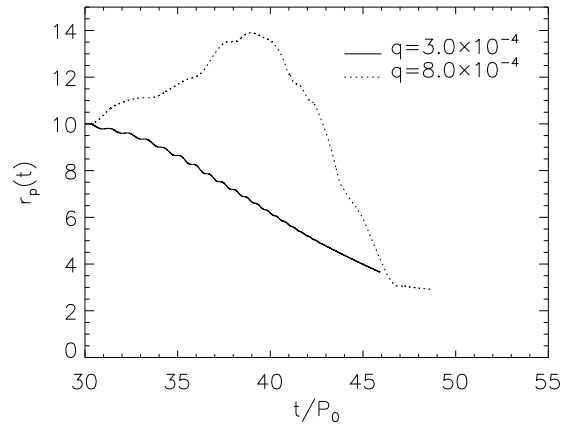}
\caption{Inward migration of sub-Jovian mass planets in a
  self-gravitating disc. The case with $q=3\times10^{-4}$
  does not develop edge modes at all. 
  \label{migin}} 
\end{figure}

We find rapid inward migration for planet masses less than about
Jovian ($q\lesssim10^{-3}$). Fig. \ref{migin} shows two examples with
$q=3\times10^{-4}$ and $q=8\times10^{-4}$. 

\subsubsection{$q=3\times10^{-4}$}
This case undergoes inward migration after release, and falls in
within $\sim15P_0$. This timescale is consistent with type III migration
\citep{masset03,peplinski08b}. 

\begin{figure}
\centering
\includegraphics[width=\linewidth]{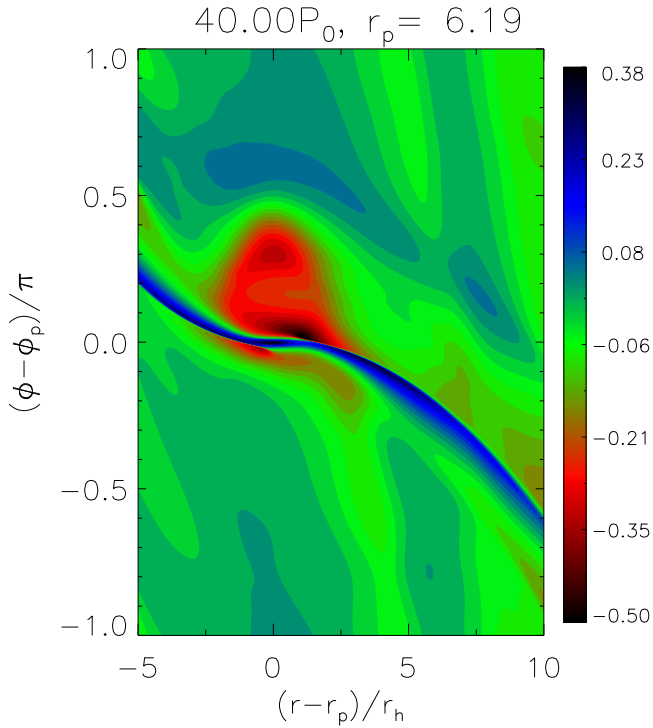}
\caption{The logarithmic surface density perturbation during inward
  migration of a planet with $q=3\times10^{-4}$. Notice the surface density 
  excess(deficit) seen behind(ahead of) the azimuthal position of the planet. 
  Such an asymmetry is characteristic of inward type III migration.} 
  \label{asymm}
\end{figure}

The occurrence of type III migration for $q=3\times10^{-4}$ is further
evidenced in Fig. \ref{asymm} where we plot the surface density during
the inward migration. The front-back density asymmetry reflects strong  
co-orbital negative torques originating from fluid crossing the
planet's orbital radius by executing outward horseshoe turns from the
inner disc ($r<r_p$) to the outer disc ($r>r_p$) behind the planet
($\phi<\phi_p$). 

Note the inward migration timescale for $q=3\times10^{-4}$ is
comparable to that needed for edge modes to develop when the planet
was held on fixed orbit. However, the planet migrates significantly
before a sufficiently deep gap can be opened to induce
instability. Therefore edge modes are irrelevant in this case. 

\subsubsection{$q=8\times10^{-4}$}

The orbital evolution for $q=8\times10^{-4}$ displays a more
complicated behaviour. In this case we found a disturbance at the
outer gap edge has already developed at planet release. 
It caused the planet to further interact with the outer gap edge,
scattering fluid inward, resulting in rapid outward migration on a
timescale of $\sim10P_0$. 

\begin{figure}
\centering
\includegraphics[scale=0.27]{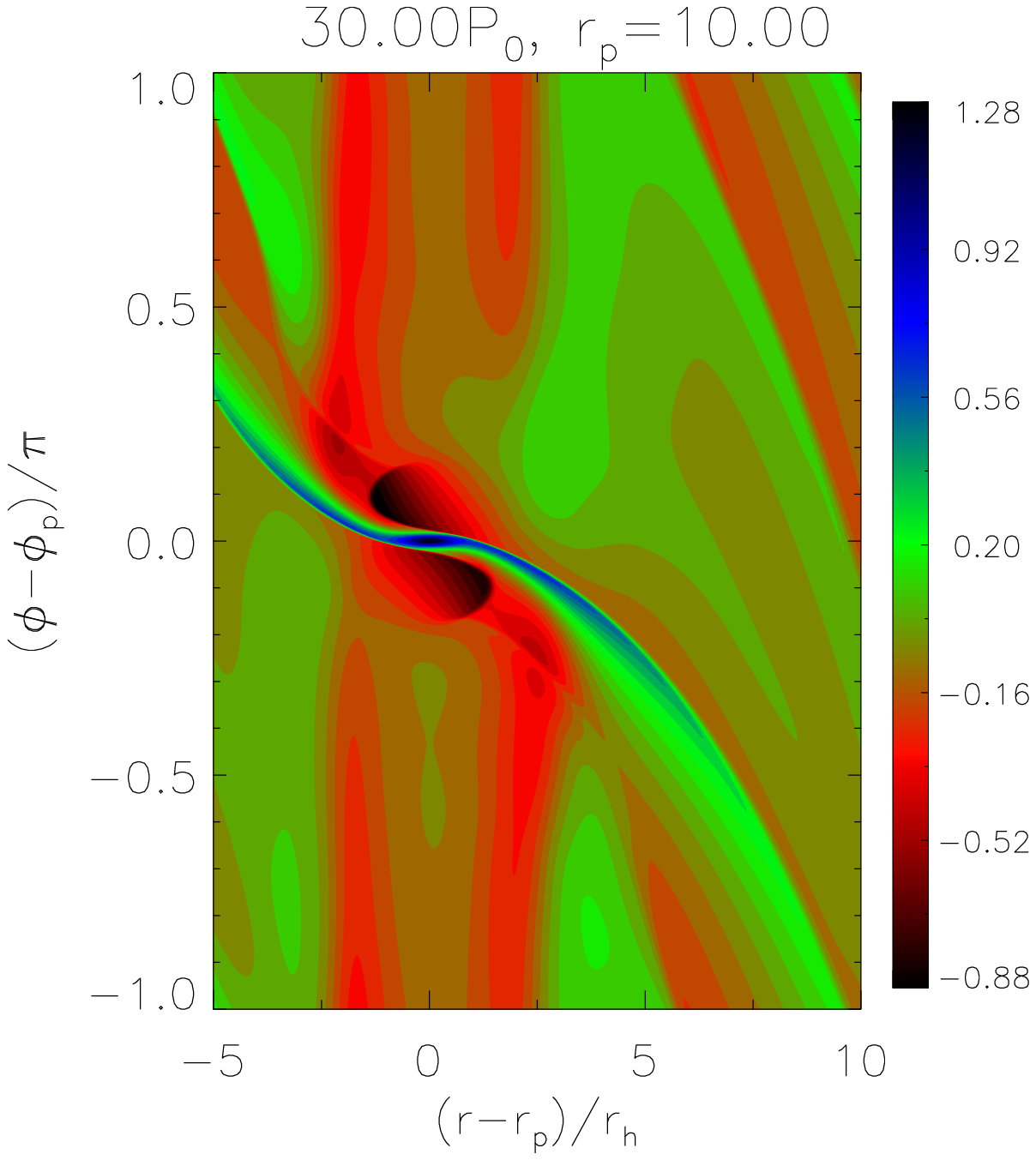}\includegraphics[scale=0.27,clip=true,trim=2.21cm
  0cm 0cm
0cm]{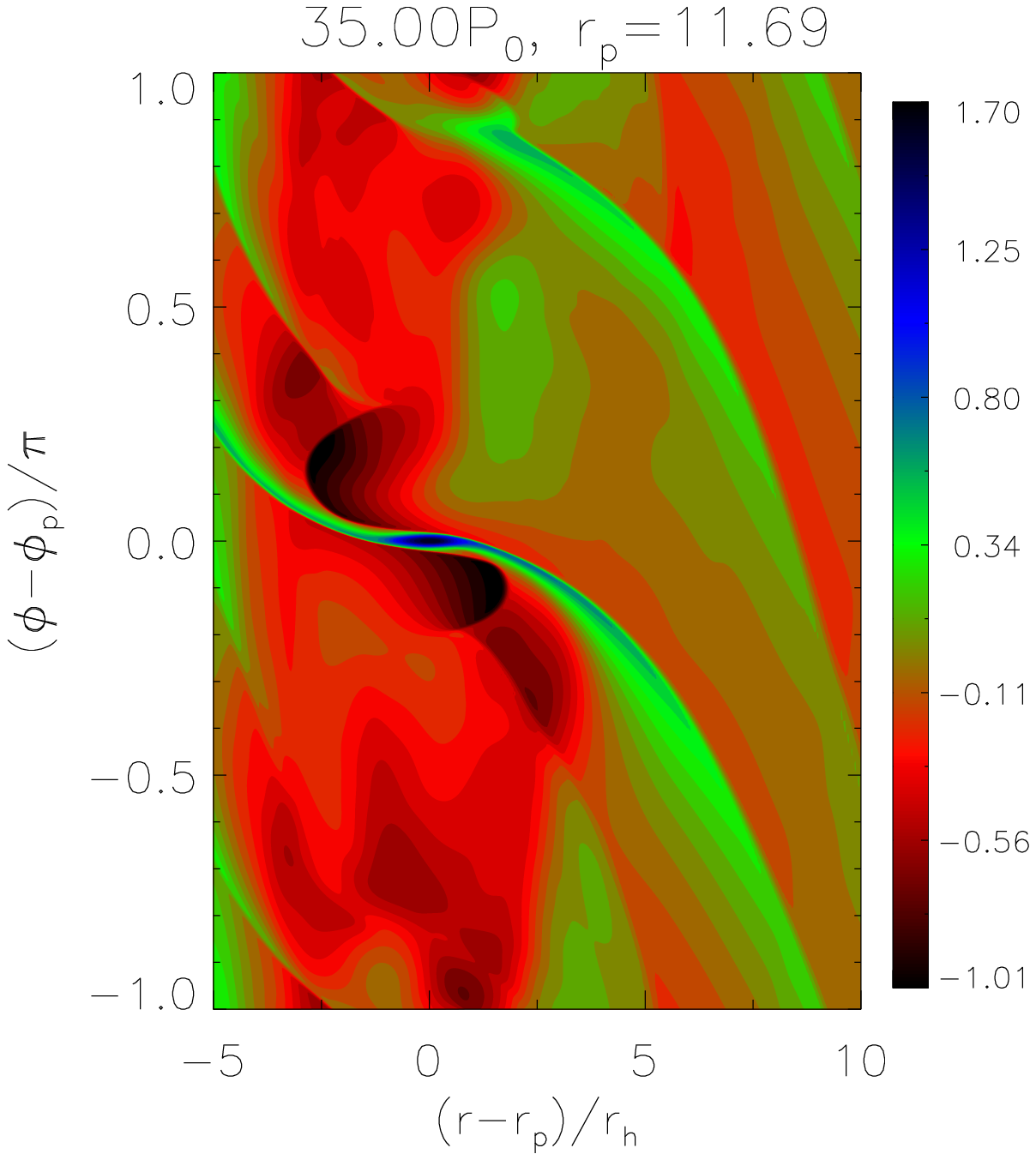}\includegraphics[scale=0.27,clip=true,trim=2.21cm
  0cm 0cm 0cm]{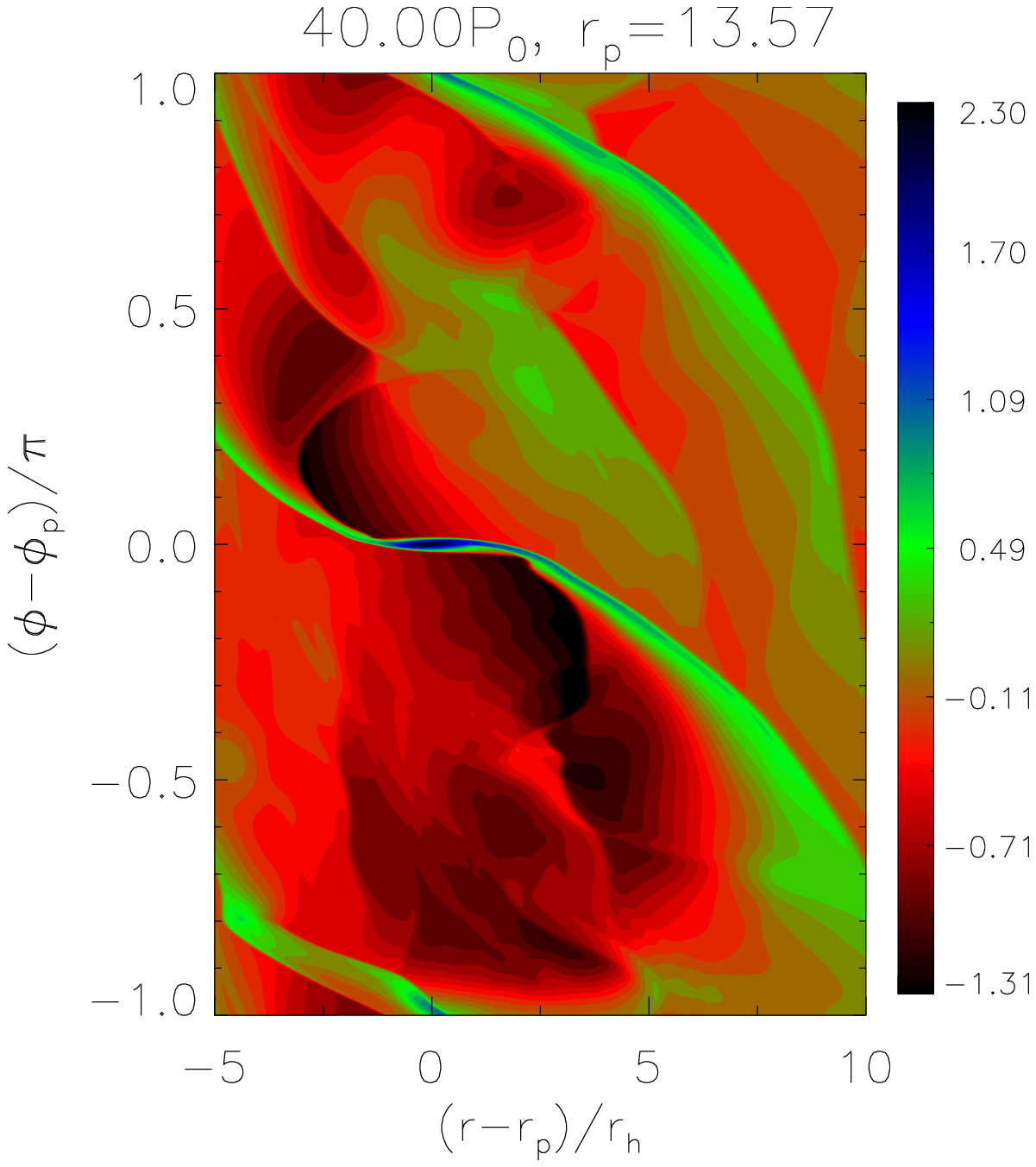} 
\caption{Initial outward migration in the case with
  $q=8\times10^{-4}$. The logarithmic surface density perturbation is
  shown. A disturbance at the outer gap edge develops
  early on, which applies a positive co-orbital torque on the planet
  (through material executing inward horseshoe turns ahead of the
  planet). 
} \label{q0.8}
\end{figure}

Fig. \ref{q0.8} shows snapshots during the initial outward
migration. The increase in $r_p$ during this phase is $\sim 6$ Hill
radii (measured at the initial orbital radius of $r=10$). So the
planet is `kicked out' of its original co-orbital region.  


However, we find the planet eventually undergoes inward type III
migration after reaching a maximum orbital radius of $r\simeq1.4r_{p0}$. This may be 
because the edge disturbance responsible for the initial kick becomes
ineffective as the planet has migrated out of its gap. 
Negative Lindblad torques then initiated inward type III migration. The
outward-inward migration seen here is similar to that observed in
\cite{lin11b} where a planet scatters off an edge mode spiral arm.    

\subsection{Outward migration of Jovian planets}   
\begin{figure}
\centering
\includegraphics[width=\linewidth]{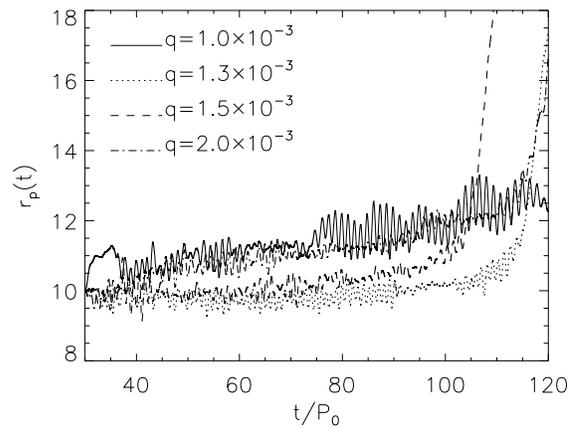}
\caption{Outward migration of massive planets 
  which open gaps that are unstable to edge modes. 
  \label{mig}} 
\end{figure}

We find planet masses  with $q \gtrsim 10^{-3}$ induced sufficiently strong
edge instabilities early on to counter-act the tendency for 
inward migration (Specifically to that of inward type III migration as
observed for $q=3\times10^{-4}$.). Several examples of such cases
are shown in Fig. \ref{mig}. The case with $q=2\times10^{-3}$ was 
considered in \citetalias{lin12} and is reproduced here for reference
(with a longer simulation time). Simulations with
$10^3q=1.3,\,1.5$  were performed to explore the possibility of a
torque balance to achieve zero net migration. 
We see from the figure that this is not
possible, and we discuss this issue in more detail later. 

Fig. \ref{mig} shows that, for $q\geq0.0013$, orbital migration 
proceeds, on average, outward more rapidly with increasing planetary mass. 
The run with $q=10^{-3}$ does not fit this trend, however. We examine
these cases separately below. 

\subsubsection{$q=10^{-3}$}
\begin{figure}
  \centering
  \includegraphics[scale=0.39]{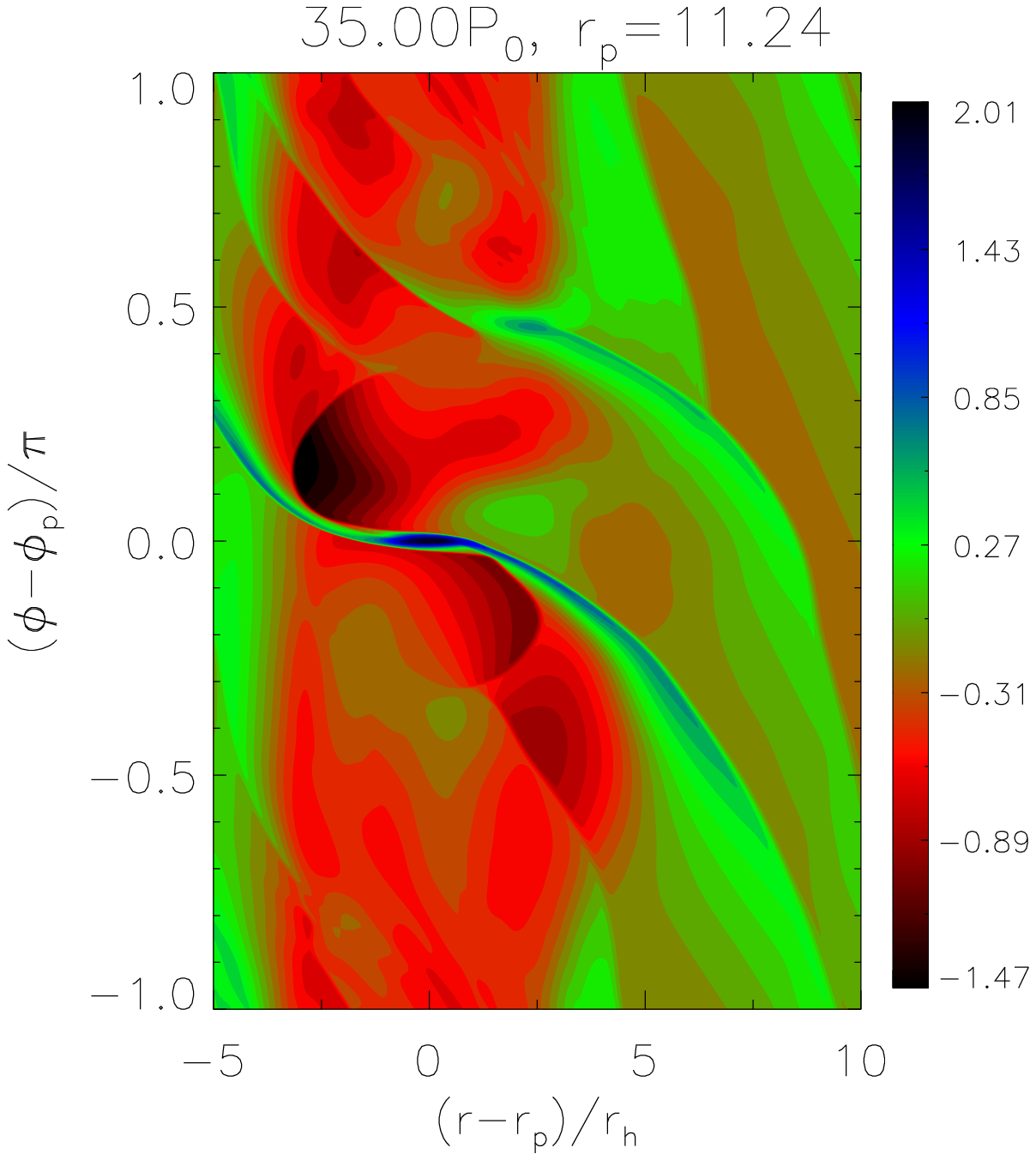}\includegraphics[scale=0.39,clip=true,trim=2.21cm 
    0cm 0cm 0cm]{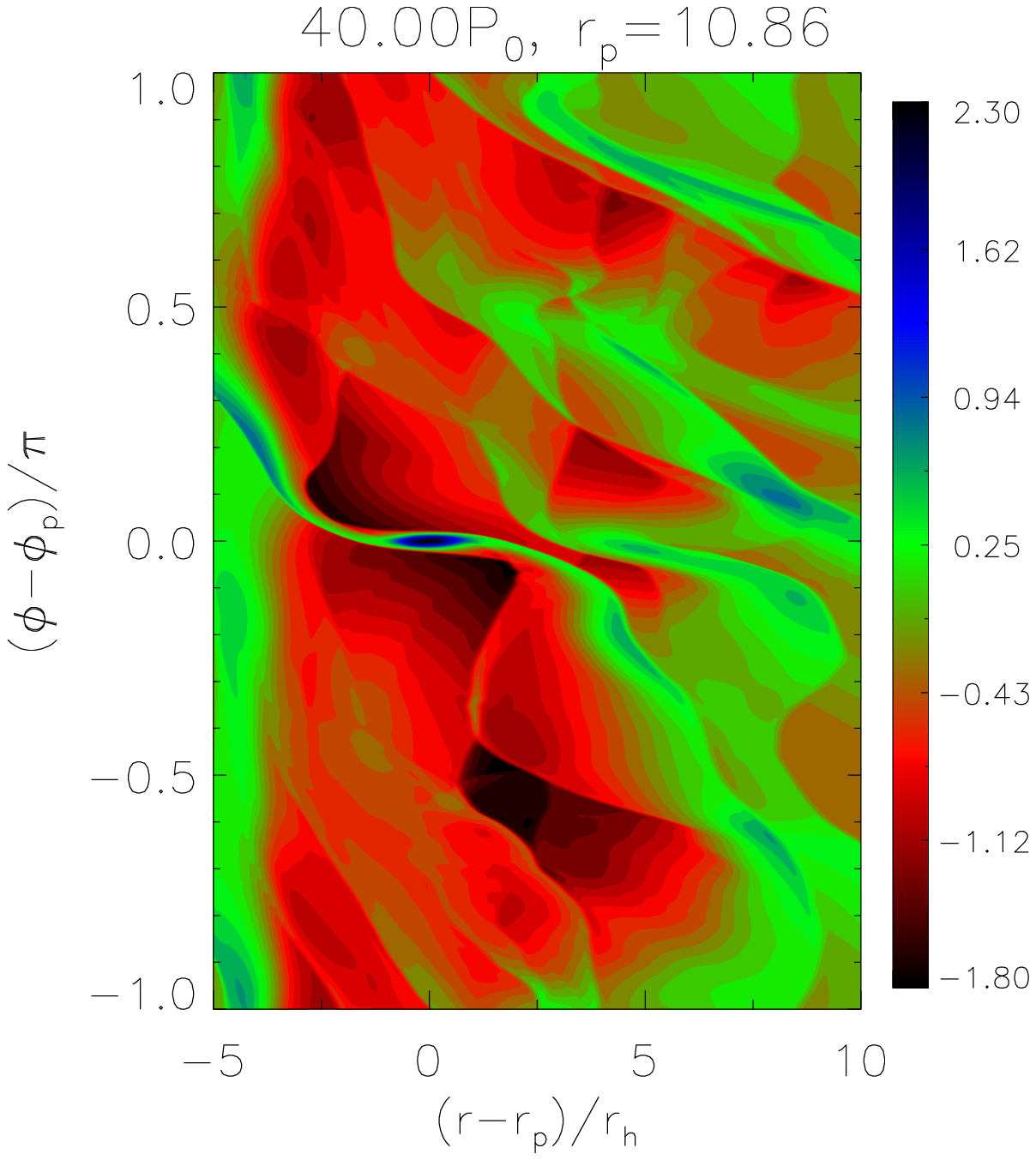} 
  \caption{Initial outward migration of the planet with $q=0.001$
    (left) and the disruption of the outer disc afterwards
    (right). 
  } \label{q1.0}
\end{figure}

In this case the planet experiences an initial kick similar
to $q=0.0008$ in the previous section. However, 
with $q=0.001$ the initial outward migration in $t=[30,35]P_0$
corresponds to about 1.4 Hill radii (at $r=10$), implying the
planet remains in its gap afterwards. This is shown in the left panel
of Fig. \ref{q1.0}. The smaller kick may be due to the increased
planet inertia: the fluid mass within the planet's Hill sphere $M_h$ plus
$M_p$ is $0.0017M_*$ for $q=0.001$ and $0.0012M_*$ for $q=0.0008$ (at
planet release).  

The planet, having remained inside the gap, does not experience the
rapid inward type III migration observed previously. However, we find
the outer disc ($r>r_p$) became highly unstable and dynamic (including 
transient clumps), as shown in the right panel of Fig. \ref{q1.0}. 
We suspect this is attributable to the passage of the spiral arm upstream of
the planet (at $r=r_p+2.5r_h$ and $\phi-\phi_p=0.5\pi$) across the
planet-induced outer wake.  

We did not identify large-scale edge mode spiral arms to persist
after the initial outward kick, because it led to a large increase in
the effective planet mass ($M_h\sim 0.0037M_*$ by $t=40P_0$), which 
strongly perturbs the outer disc directly through the planet-induced
wake (becoming unstable itself). The planet nevertheless migrates
outwards secularly because the inner gap edge is on average closer to
the planet than the outer gap edge. 

\subsubsection{$q\geq1.3\times10^{-3}$}\label{massive}
For these cases, Fig. \ref{mig} displays outward migration on two
distinct timescales: a relatively slow phase over several 10's of
$P_0$ followed by an outward kick towards the end of the simulation.  

The mechanism responsible for the first phase was described in
\citetalias{lin12}. In summary, the passage of outer edge mode spirals
by the planet supplies material to execute inward horseshoe turns upstream of
the planet. This applies a positive co-orbital torque, and overcomes
the negative differential Lindblad torque on average. We find the 
net outward migration during this phase is roughly linear in 
time. 
As Fig. \ref{mig} shows, decreasing $q$ and hence instability 
strength, the migration rate is reduced during this phase. This is 
despite the expectation that lowering $q$ should permit higher surface 
densities within the gap.

The second phase of outward migration is very fast and almost
monotonic. For $q=0.002$, $r_p$ increases by about $30\%$ in
$t\in[110,120]P_0$. For $q=0.0013$ and $0.0015$,  $r_p$ increases by
more than $60\%$ over a similar timescale. We will examine this phase
in more detail in the next section. The extent of outward migration
during the second phase is much larger than the initial  
kicks observed in the previous section with smaller $q$. 
We caution that the simulation should not be trusted close to and after 
$\mathrm{max}(r_p)$ has been attained (i.e. after $t\sim 120P_{0}$ for $q=0.0013$), because the planet's co-orbital 
region approaches the outer disc boundary.

\begin{figure}
\centering
\includegraphics[width=\linewidth]{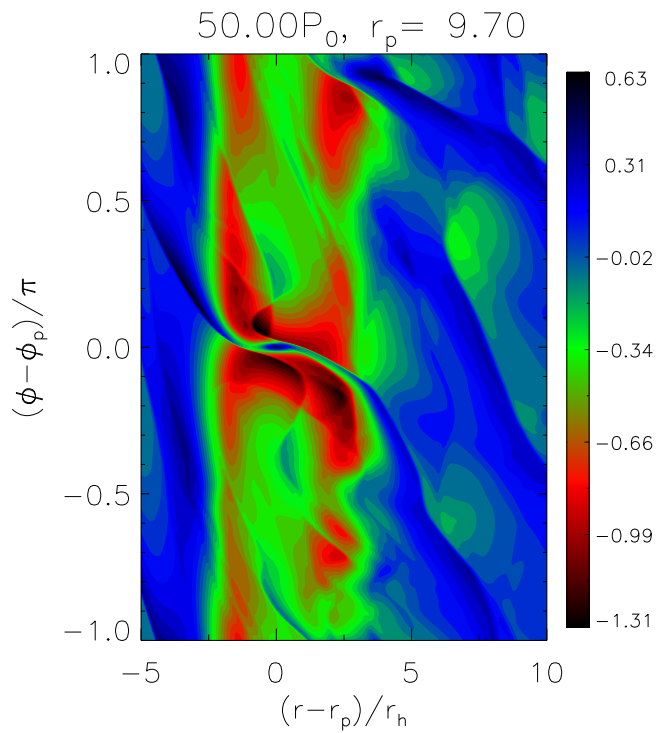}
\caption{Simulation with a freely migrating planet with mass
  $q=0.0013$. 
  This  snapshot is taken during
  the first phase of its migration history when there is a rough
  balance between the positive torques provided by the outer edge mode
  spirals and the negative Lindblad torques from the planet-induced
  wakes (see \S\ref{massive}). However, the planet eventually
  interacts with the outer disc and is scattered outwards. 
} \label{q1.3}
\end{figure}

It is interesting to note that, had we not simulated the case with
$q=0.0013$ beyond $t\simeq 100P_0$, the first phase would suggest
little net migration. A snapshot is shown in Fig. \ref{q1.3}. 
However, it still enters the second phase due to
interaction with the gap edge as discussed next. Thus, it appears
unlikely that a state may be reached in which positive torques
due to edge modes balance against the negative differential Lindblad
torques, keeping the planet at a fixed orbital radius 
\citepalias[which was hypothesized in][]{lin12}.   

\section{Rapid outward type III migration induced by edge
  modes} \label{fiducial}
A common feature in simulations with a freely migrating
planet with mass $q\geq0.0013$ is the second phase of rapid outward
migration. 

\begin{figure}
  \centering
  \includegraphics[width=\linewidth]{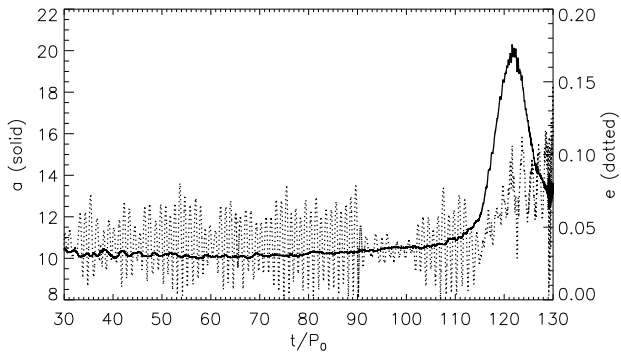}
  \caption{Evolution of semi-major axis $a$ (left, solid) and
    eccentricity $e$ (right, dotted) for the case with $q=0.0013$.  
    \label{q1.3_a_e}} 
\end{figure}

For reference, we show the orbital evolution for $q=0.0013$ in
Fig. \ref{q1.3_a_e}. The semi-major axis $a$ and eccentricity $e$ were
calculated assuming a Keplerian orbit about the central star. The
orbit remains fairly circular ($e\lesssim0.1$) and maintains
$a\simeq10$ for $t\lesssim80P_0$, but experiences a
strong outward kick at $t\sim 100P_0$. This kick appears almost
spontaneously. To understand its origin, we examine the fiducial case
$q=0.0013$ in more detail in the proceeding section. This 
simulation was also performed at a lower resolution ($N_r\times N_\phi=512\times1024$), which
exhibited the same qualitative behaviour as the high resolution ($N_r\times N_\phi = 1024\times2048$) 
results discussed here.


We remark that the characteristic kick may also occur in 
less massive discs. Specifically, we performed low resolution
simulations of a $q=0.0013$ planet migrating in discs with $Q_{k0}=1.7,2.0$. 
In a disc with $Q_{k0}=1.7$ the planet gets kicked outwards at $t\sim110 
P_{0}$ similar to the fiducial case ($Q_{k0}=1.5$ at low resolution) but at a slightly later stage in the planet's
evolution. In a disc with $Q_{k0}=2.0$ the  planet remains at $r_{p}\sim10$ for the length of the simulation 
($t\in[30,150]P_{0}$). This is similar to the behaviour of our fiducial
case prior to the characteristic kick, which suggests the $q=0.0013$
planet in the $Q_{k0}=2.0$ disc may eventually experience the same 
kick evident in Fig. \ref{q1.3_a_e}.  



\subsection{Interaction with the outer gap edge} \label{5.1}
We first show that the kick is associated with the planet migrating 
into the outer gap edge. We define the dimensionless outer 
gap width $w_\mathrm{,out}$ as 

\begin{align}
  w_\mathrm{out}(t)=\frac{(r_\mathrm{out}-r_{p})}{r_{h}},
\end{align}
where we recall $r_\mathrm{out}$ is the radius of the outer gap edge 
(\S\ref{gapsect}). 
This definition is only valid when the planet resides in the gap
($w_\mathrm{out} > 0$). 

\begin{figure}
\centering
\includegraphics[scale=0.5,clip=true,trim=0cm 0cm 0cm 0cm]{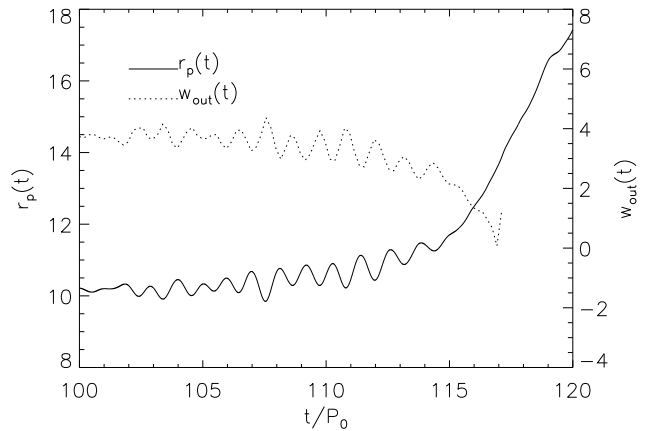}
\caption{The outer gap width (right, dotted) from $t=100P_{0}$ until the planet
 receives a kick from the unstable gap edge. The planet's orbital radius
  $r_p$ is also shown (left, solid). Note that the gap width is
  undefined when the planet no longer resides in an annular gap, here
  observed to occur after $t=117P_0$.  
  \label{1d3mig}} 
\end{figure}

The outer gap width leading up to the kick is plotted in 
Fig. \ref{1d3mig}, along with $r_p(t)$. 
For $t\lesssim 110P_0$, $w_\mathrm{out}$ oscillates on orbital
time-scales with an average value of $4$. The oscillations are 
due to the periodic passage of non-axisymmetric over-densities
associated with the outer gap edge mode by the planet.

There is a tendency for inward migration due to negative Lindblad torques
except when an edge mode spiral is approaching the planet from
upstream, by supplying some material to execute inward horseshoe
turns. It is evident from the outward migration shown in Fig. \ref{1d3mig} that the positive
torques supplied by the edge mode spiral overcome negative torques on
average. The planet migrates outward with respect to the outer gap
edge, so $w_\mathrm{out}$ decreases.    

Notice $w_\mathrm{out}$ stops oscillating after
$t\simeq115P_0$, and the kick commences.  At this point
$w_\mathrm{out}\simeq 2$. That is, the kick occurs when the outer 
gap edge enters the co-orbital region of the planet. (Recall
  the co-orbital region of a giant planet is 
  $|r-r_p|\lesssim 2.5r_h$.) 
Afterwards $r_p$ rapidly increases and $w_\mathrm{out}\to0$,
implying the planet has exited the gap. 

\begin{figure*}
\centering
\includegraphics[scale=0.55,clip=true,trim=0cm 1.8cm 0cm 0cm]{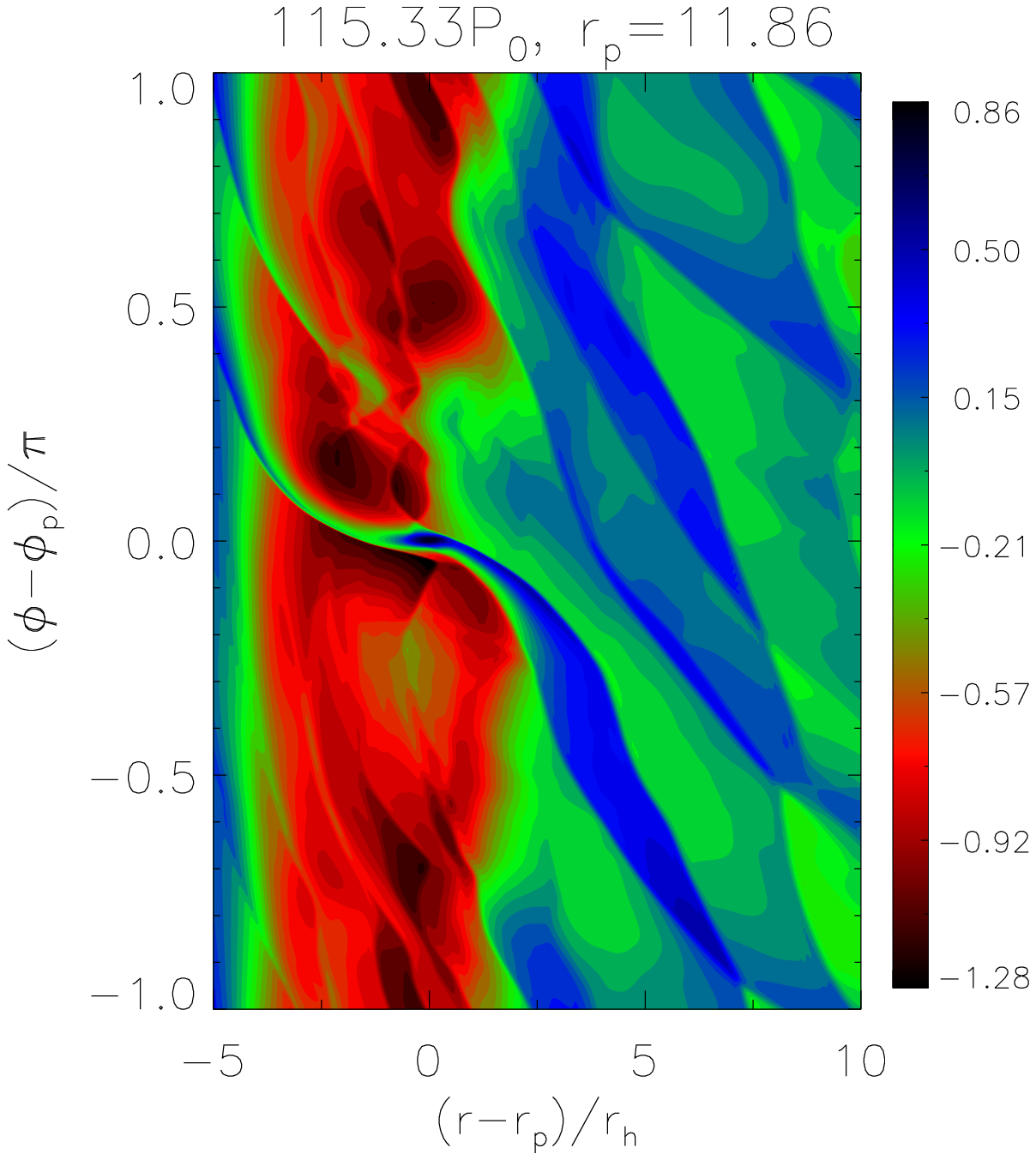}
\includegraphics[scale=0.55,clip=true,trim=2.2cm 1.8cm 0cm 0cm]{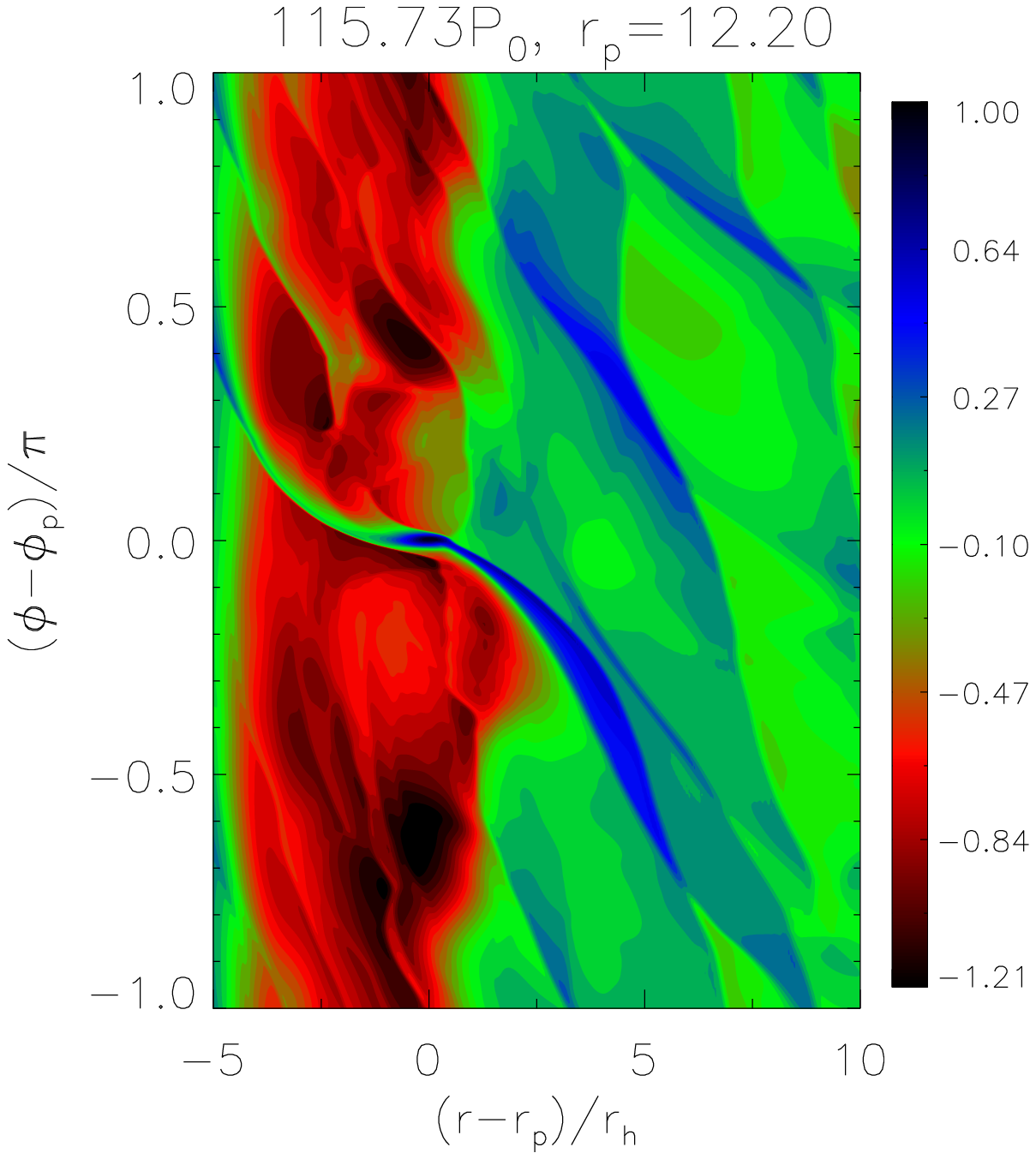}
\includegraphics[scale=0.55,clip=true,trim=2.2cm 1.8cm 0cm 0cm]{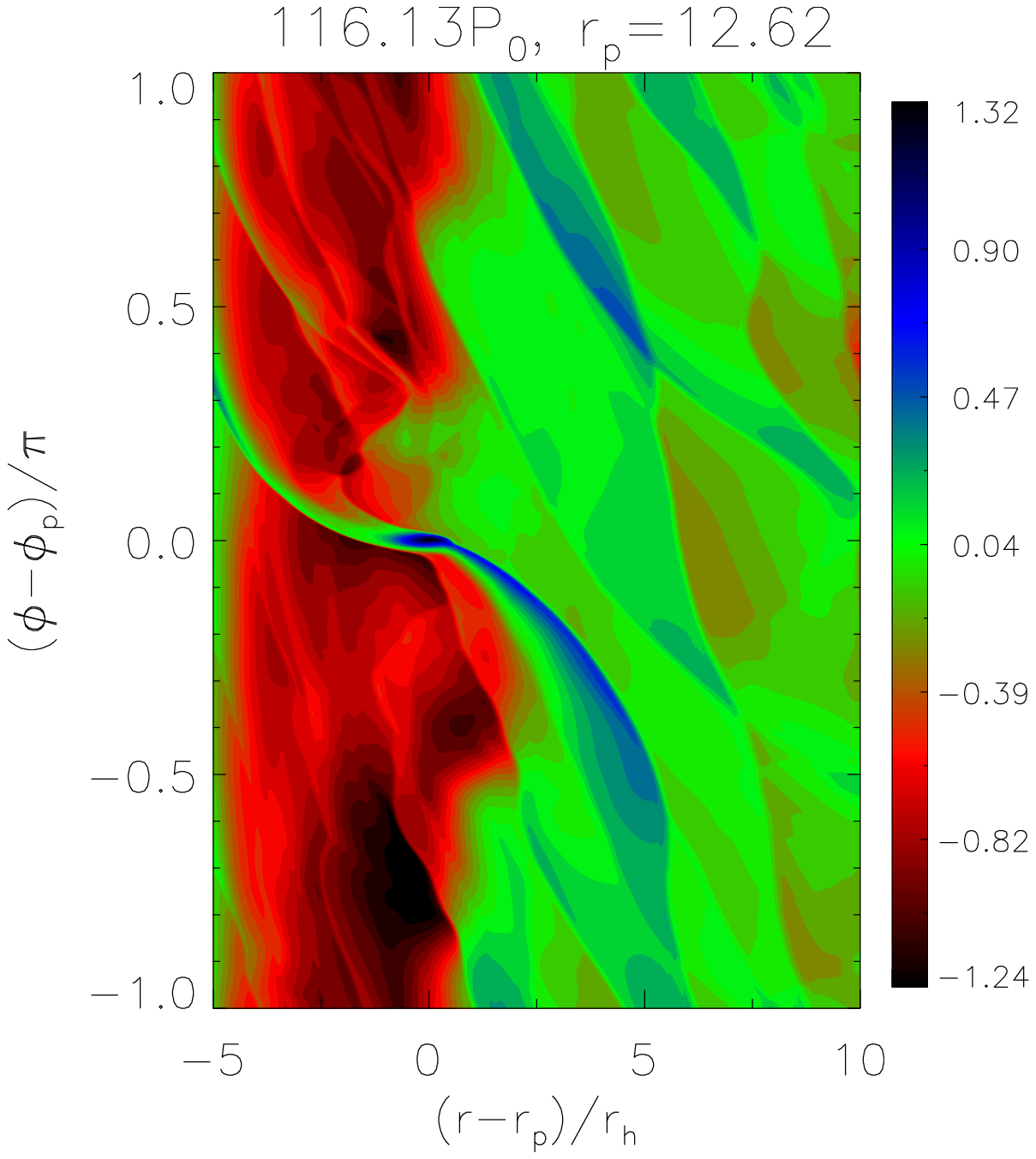} \\
\includegraphics[scale=0.55,clip=true,trim=0cm 0cm 0cm 0cm]{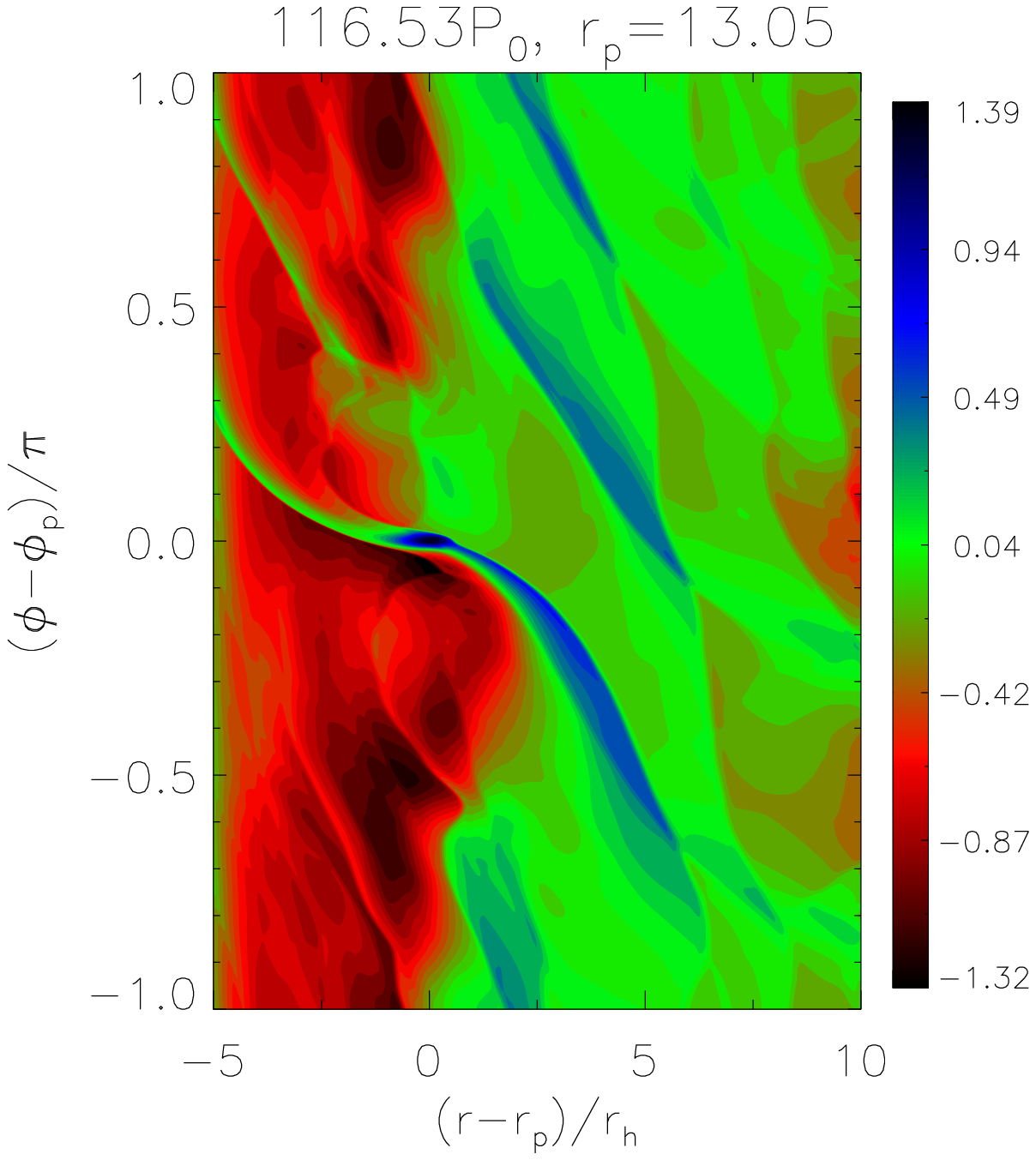} 
\includegraphics[scale=0.55,clip=true,trim=2.2cm 0cm 0cm 0cm]{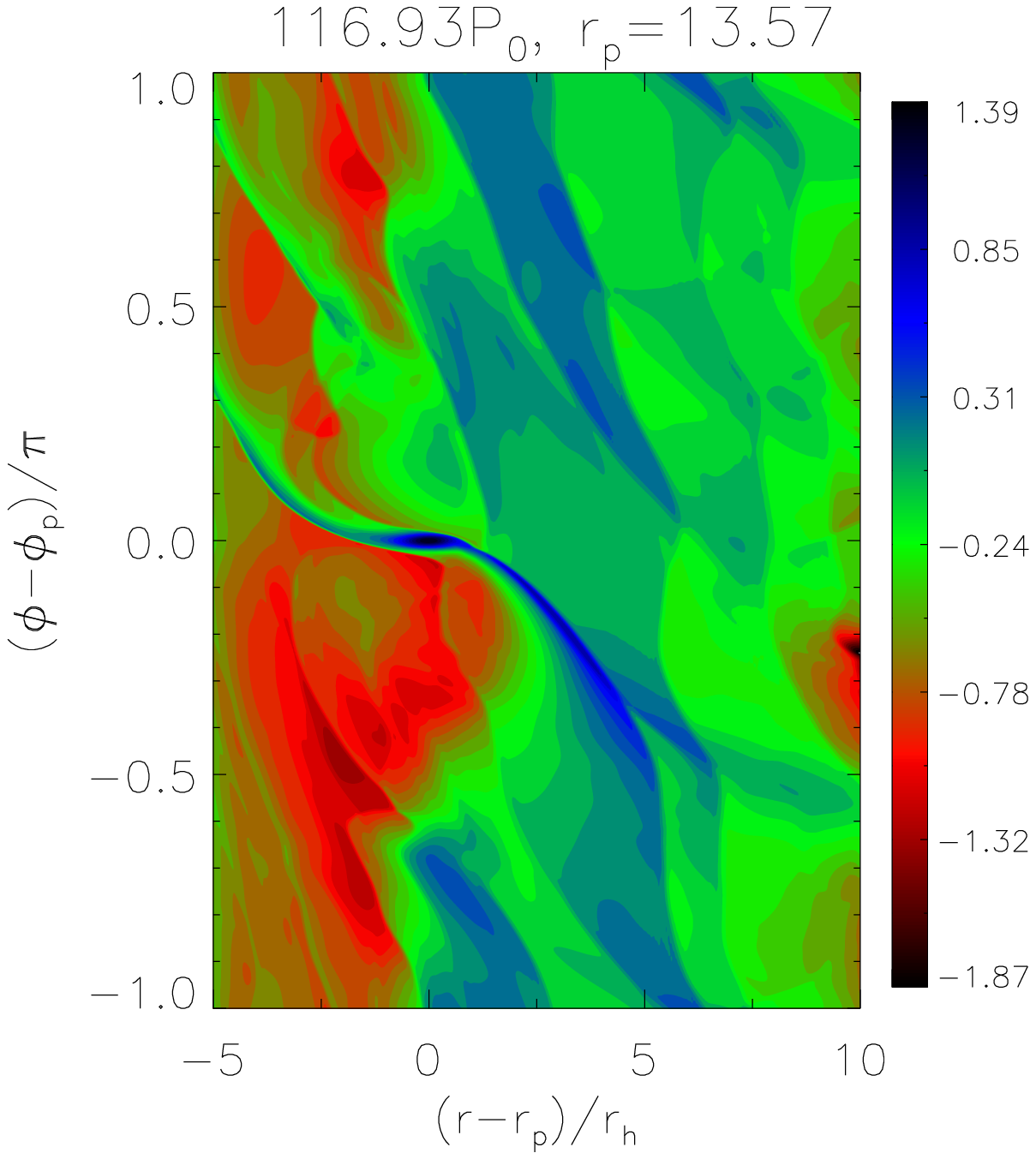}
\includegraphics[scale=0.55,clip=true,trim=2.2cm 0cm 0cm 0cm]{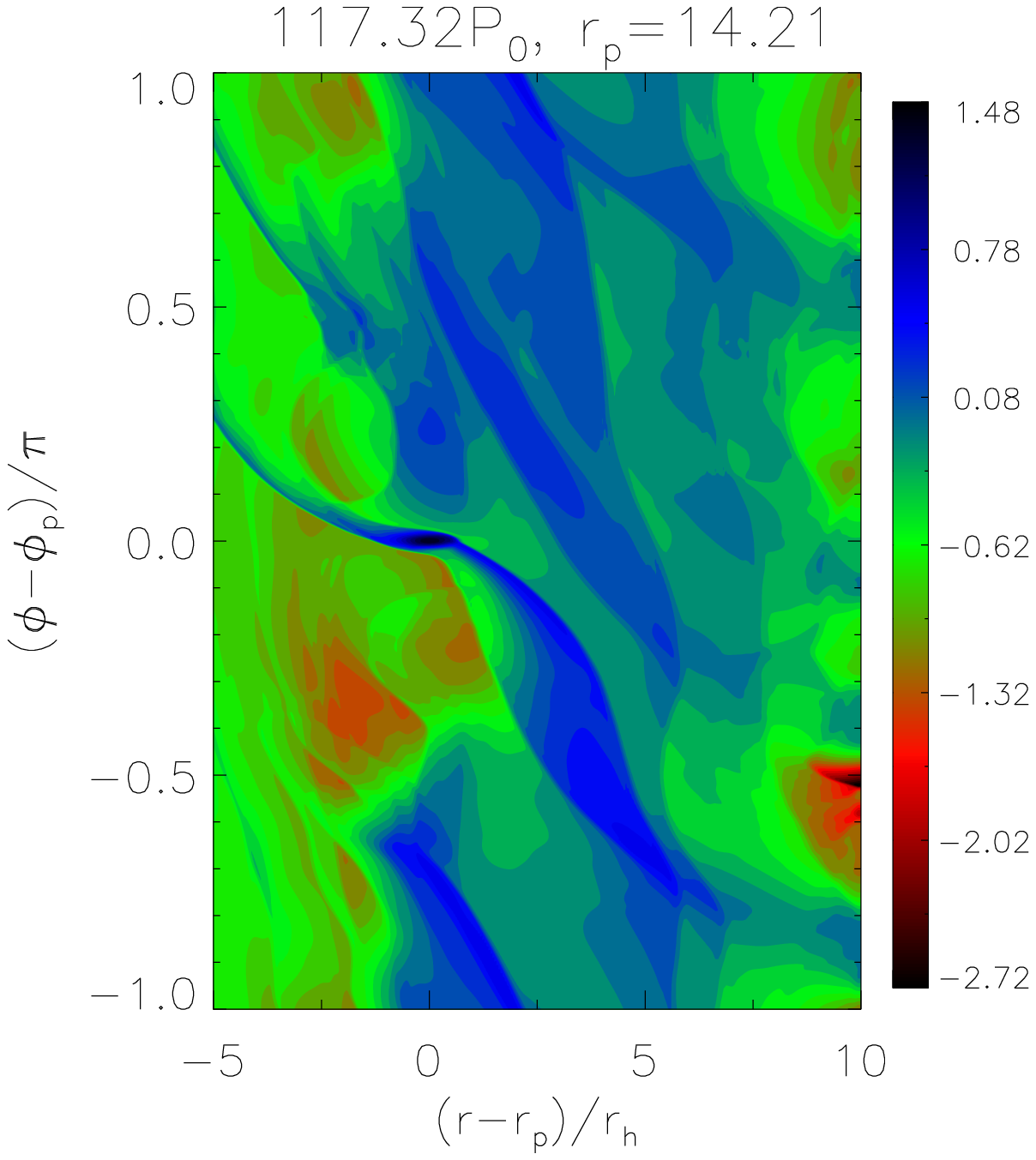} 
\caption{Outward type III migration triggered by an edge mode spiral in
  the case with $q=0.0013$. There is relatively little
  migration for $t\lesssim100P_0$, but eventually the planet scatters
  off an edge mode spiral into the outer disc. The logarithmic surface density
  perturbation is shown.
  \label{kick}} 
\end{figure*}

Fig. \ref{kick} shows the interaction between the planet and the outer
gap edge during this kick. In this case, rather than passing by the
planet, the bulk of the edge mode over-density executes an inward 
horseshoe turn upstream of the planet, i.e. the planet 
scatters fluid comprising the outer gap edge inwards. The resulting
azimuthal surface density asymmetry about the planet is that which
characterizes outward type III orbital migration
\citep{peplinski08b}. An under-dense(over-dense) region behind(ahead of)
the planet implies a strong positive co-orbital
torque. This configuration persists until the planet migrates
close to the outer disc boundary. We conclude that the outer edge
mode spiral arms act as a natural `trigger' for outward type III migration.   

\subsection{Growth of effective planet mass} \label{effmass} 
We also observe a significant increase in the effective planet mass
after the planet enters outward type III migration. We define
$M_\epsilon$ as the fluid mass contained within a   
radius $\epsilon r_h$ of the planet. For $\epsilon = 1$, $M_\epsilon$
equals $M_h$ defined previously. Note that $M_\epsilon$ includes fluid
gravitationally bound to the planet \emph{and} orbit-crossing fluid,
the latter being responsible for type III migration.   

\begin{figure}
\centering
\includegraphics[width=\linewidth]{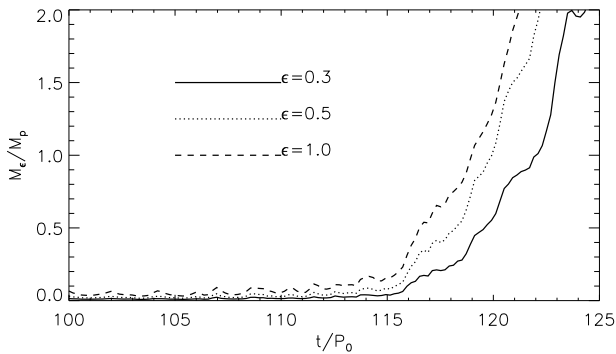}
\caption{Evolution of fluid mass contained within a fraction
  $\epsilon$ of the planet's Hill radius in the case with
  $q=0.0013$. The time interval corresponds to the planet just prior
  to and after the interaction with an edge mode spiral arm which
  scatters it outwards.  
  \label{effective_mass}} 
\end{figure}

In Fig. \ref{effective_mass} we plot $M_\epsilon/M_p$ for $\epsilon =
0.3,\,0.5,$ and $1.0$. For $t\lesssim110P_0$ there is negligible mass
contained within the planet's Hill radius. This mass rapidly
increases as the planet interacts with the edge mode spiral arm
($t\simeq115P_0$). Notice $M_{\epsilon\geq0.5}$ increases more rapidly
than $M_{\epsilon=0.3}$, suggesting a significant flux of
orbit-crossing fluid.    

The planet migrates to a maximum orbital radius $\mathrm{max}(r_p)\simeq
18$ at $t\simeq 120P_0$. At this point $M_{\epsilon=0.3}\simeq
0.6M_p$ and the fluid within the Hill radius exceeds the planet mass,
$M_h\simeq 1.3M_p$. The rapid increase in planet inertia may attribute to
stopping type III migration \citep{peplinski08c}. However, since the
planet's co-orbital region approaches the disc boundary ($r_p+2.5r_h
\gtrsim 0.85r_\mathrm{out}$), boundary conditions may come into effect 
(e.g., the lack of disc mass exterior to the planet to 
sustain type III migration). 

Nevertheless, we expect the effective planet mass to generally
increase if it undergoes outward type III migration because the Hill
radius $r_h$ scales with orbital radius. In this sense, edge modes can
indirectly increase the effective planet mass. 


\section{Summary and discussion}\label{summary}

We have performed hydrodynamic simulations of gap-opening
satellites (planets) in self-gravitating discs. Our aim was
to examine the role of planet mass on the development of
gravitational instability associated with the gap, and its
subsequent effect on orbital migration.  

We first considered simulations where the planet was held on a fixed
orbit. Outer gap edge modes developed for planet-to-star mass ratios
$q\in[0.3,2.0]\times10^{-3}$. Despite being a gravitational
instability, we find edge modes developed earlier and are stronger
with increasing $q$, which corresponds to deeper gaps with lower surface
density. This is because edge modes are fundamentally associated with
potential vorticity maxima resulting from planet-induced spiral shocks, which
become sharper with increasing planet mass \citep{lin10}. We also
find disc-on-planet torques typically become more positive with
increasing $q$. This is consistent with positive torques being
supplied by the outer gap edge mode spirals \citepalias{lin12}.   

However, in simulations where orbital migration is allowed, we found
edge modes were only relevant for planet masses that opened an unstable
gap early on. This is because the condition required for edge modes to
develop --- a massive disc --- also favours type III migration, which
operates on dynamical timescales. We found for $q=0.0003$, the planet immediately
underwent inward type III migration, having no time to open a gap. 

This initial inward type III migration could be avoided due to
  initial conditions \citep{artymowicz04}. \cite{peplinski08c} demonstrated outward type
  III migration of giant planets when they are initially placed at the
  edge of an inner cavity. Thus, an appropriate 
  choice of initial surface density profile may prevent a Saturnian mass
  planet ($q=0.0003$) to undergo immediate rapid inward migration as seen in our
  simulations, and allow it to remain at approximately the
  same orbital radius for at least $\sim 20$ orbits. Then edge modes will become
  relevant, because our fixed-orbit simulations indicate that even the
  partial gap opened by $q=0.0003$ is unstable.

For $q\geq0.0013$ we find net outward migration with an increasing rate
with planet mass and therefore gap instability. Although we
found it was possible, by decreasing $q$, to achieve almost zero net
migration, this phase only lasted a few $10$'s of orbital periods. The
planet eventually interacted with an outer gap edge mode spiral and
underwent outward type III migration. We conclude that the
scenario hypothesized by \citetalias{lin12} --- inwards type II
migration begin balanced by the positive torques due to outer gap edge
mode spirals --- is a configuration that is unlikely to persist beyond
a few tens of orbital periods.

\subsection{Relation to previous studies}
Recent hydrodynamic simulations which focus on the orbital migration of 
giant planets in massive discs have been motivated by the possibility
of planet formation through gravitational instability of an initially
unstructured disc \citep{boss05,boss13,baruteau11,michael11}. Such disc
models are already gravitationally unstable without a planet. These
studies employed models in which the planet does not significantly
perturb the gravito-turbulent disc, and no gap is formed. By contrast,
the gravitational instability in our disc model is \emph{caused by} the 
planet indirectly through gap formation. 

Detailed disc fragmentation simulations have shown
that, while most clumps formed through gravitational instability are
lost from the system (e.g. by rapid inward migration), in some cases
it was possible to form a gap-opening clump \citep{voro10,voro13,zhu12}. 
These authors did not specifically examine gap stability, but we note
several interesting results that may hint the presence of edge modes. 
 
In \cite{voro10}, a clump is seen to migrate outward. 
Just prior to this outward migration, their Fig. 1 shows a spiral arm
upstream to the planet, and is situated at a radial boundary between
low/high disc surface density (but a gap is not yet
well-defined). This spiral arm may have contributed to a positive
 torque on the clump. Their Fig. 1 also suggest an eccentric
gap. This is probably due to clump-disc interaction
\citep{papaloizou01,kley06,dunhill13}, but eccentric gaps have also
been observed in disc-planet simulations where edge modes develop and
saturate \citep[][Fig. 12]{lin11b}.  

\cite{voro13} improved upon \cite{voro10} by including detailed
thermodynamics. No edge mode spirals are visible from their plots. 
However, they found that in all cases where a gap-opening clump is
formed, the clump migrates outward. The origin of the required
positive torque was not identified. It is conceivable that destruction
of the outer gap edge, perhaps due to the edge spiral instability,
could result in the clump being on average closer to the inner gap
edge than the outer gap edge. This will lead to outwards type II
migration.     

\cite{zhu12} also simulated the fragmentation of massive discs with
realistic thermodynamics. In one case where a gap-opening 
clump formed, further clump formation was observed at the gap
edge. The gap-opening clump first migrated outward, but
ultimately falls in. This outward migration may be due to the
interaction with the unstable gap edge, similar to our simulations.  

\subsection{Giant planets on wide orbits} 
Our results have important implications for some models seeking to
explain the observation of giant planets on wide orbits. Such examples
include the 4 giant planets orbiting HR 8799 between $15$---$68$AU
\citep{marios08,marios10}, Fomalhaut b at $115$AU from its star
\citep{kalas08,currie12}, and AB Pic b at $260$AU\citep{chauvin05}.  

Disc fragmentation has been proposed as an in situ mechanism to form
long-period giant planets \citep{dodson09, boss11,
  voro13}. \citeauthor{voro13} showed that this was possible if the 
clump opened a gap to avoid rapid inward migration \citep{baruteau11}.
Then gravitational gap stability becomes an issue that should be 
addressed. Our results indicate an unstable gap can help to prevent
inward migration, but there is the danger that it may scatter the
planet if edge modes persist. The gravitational edge
  instability is therefore a potential threat to clump survival; in
  addition to other known difficulties with the disc  
  fragmentation model \citep[see][]{kratter10}.

On the other hand, 
our simulations reveal a way for a single giant planet to migrate 
outward, by opening a gravitationally unstable gap and
letting it trigger rapid outward type III migration. Our 
simulations indicate that such outward migration will increase the
planet's effective mass, which may contribute to a 
circumplanetary disc. Indeed, circumplanetary discs associated with
planets on wide orbits have been observed \citep{bowler11}. In our
models it is not clear whether the rapid increase in effective planet
mass discussed in $\S$ \ref{effmass} could slow down or halt the triggered
rapid outward type III migration. We cannot address this possibility 
because of the finite disc domain in our models. 
Conversely, type III migration is self-sustaining \citep{masset03}, 
so giant planets that underwent type III migration, initiated
by this  `trigger' mechanism, could be found 
at large orbital radii.

\subsection{Caveats and future work}\label{caveats}
Our study is subject to several caveats that should be clarified in
future work:

\emph{Initial conditions.} Our simulations with $q=0.0008$ and 
$q=0.001$ displayed very different results for orbital migration, 
despite having similar planet mass. We identified the planet to 
interact with a disturbance at the outer gap edge at planet
release. This initial kick provided a strong co-orbital torque. It can
be interpreted as a brief phase of type III migration, which is known to depend 
on initial conditions. A more extensive exploration of numerical
parameter space is needed to assess the importance of this initial kick.   
Specifically the resolution of simulations involving these
particular planet masses may be important for whether or not the initial 
interaction between the planet and a disturbance at the outer gap edge occurs.

\emph{Thermodynamics.} The locally isothermal equation of state 
implicitly assumes efficient cooling. This favours gap formation and
gravitational instabilities. We expect the gap to become
gravitationally more stable if the disc is allowed to heat up.  
Numerical simulations including an energy   
equation, which adds another parameter to the problem --- the cooling
time --- will be presented in our follow-up paper on the gravitational
stability of planet gaps in non-isothermal discs.

\emph{Disc geometry.} The thin-disc approximation is expected to be
valid for gap-opening perturbers, since their Hill radius exceeds the
disc thickness by definition. Indeed, three-dimensional (3D) simulations
carried out by \cite{lin_thesis} also revealed outward migration of a
giant planet due to a gravitationally unstable gap. For partially
gap-opening planets (e.g. the $q=0.0003$ case in our models), 2D works
less well, but our simulations indicate they nevertheless open
unstable gaps. Whether or not this remains valid in 3D, needs to be
addressed.   


\section*{acknowledgments}
This project was initiated at the CITA 2012 summer student programme. 
RC would like to thank CITA for providing funding throughout the
project and the use of the Sunnyvale computing cluster. Computations were
also performed on the GPC supercomputer at the SciNet HPC
Consortium. SciNet is funded by: the Canada Foundation for Innovation
under the auspices of Compute Canada; the Government of Ontario;
Ontario Research Fund - Research Excellence; and the University of
Toronto. The authors thank C. Matzner, K. Kratter and the
  anonymous referee for comments and suggestions.


\end{document}